\documentclass{ecai} 

\usepackage{latexsym}
\usepackage{amssymb}
\usepackage{amsmath}
\usepackage{amsthm}
\usepackage{booktabs}
\usepackage{graphicx}
\usepackage{color}

\usepackage{times}
\usepackage{soul}
\usepackage{setspace}
\usepackage{url}
\usepackage[hidelinks]{hyperref}
\usepackage[utf8]{inputenc}
\usepackage[small]{caption}
\usepackage{graphicx}
\usepackage{amsmath}
\usepackage{amsthm}
\usepackage{booktabs}
\usepackage[ruled, vlined, linesnumbered]{algorithm2e}
\usepackage[para]{footmisc}
\usepackage[switch]{lineno}
\let\oldnl\nl
\newcommand{\nonl}{\renewcommand{\nl}{\let\nl\oldnl}}
\usepackage[utf8]{inputenc} 
\usepackage[T1]{fontenc}    
\usepackage{hyperref}       
\usepackage{url}            
\usepackage{stfloats}
\usepackage{booktabs}       
\usepackage{amsfonts}       
\usepackage{nicefrac}       
\usepackage{microtype}      
\usepackage[dvipsnames]{xcolor}         
\usepackage{bbold}
\usepackage[super]{nth}
\usepackage{amsmath}
\usepackage{tabularx}
\usepackage{booktabs,arydshln}
\usepackage{adjustbox}
\usepackage{multirow}
\usepackage{makecell}
\usepackage[group-separator={,},group-minimum-digits={3}]{siunitx}
\usepackage{makecell}
\usepackage{caption}
\usepackage{subcaption}
\usepackage{xspace}
\usepackage[inline]{enumitem}
\usepackage{tikz}
\usepackage{pgfplots}
\usepackage{MnSymbol}
\usepackage{pifont}

\usepgfplotslibrary{colorbrewer}
\pgfplotsset{compat = 1.15, cycle list/Set1-8} 
\usetikzlibrary{pgfplots.statistics, pgfplots.colorbrewer} 
\usepackage{pgfplotstable}
\usepackage{filecontents}

\usepackage{booktabs}
\usepackage[normalem]{ulem}
\useunder{\uline}{\ul}{}

\newcommand{\PreserveBackslash}[1]{\let\temp=\\#1\let\\=\temp}
\newcolumntype{C}[1]{>{\PreserveBackslash\centering}p{#1}}
\newcolumntype{R}[1]{>{\PreserveBackslash\raggedleft}p{#1}}
\newcolumntype{L}[1]{>{\PreserveBackslash\raggedright}p{#1}}

\newcommand{\algrule}[1][.2pt]{\par\vskip.5\baselineskip\hrule height #1\par\vskip.5\baselineskip}

\let\oldding\ding
\renewcommand{\ding}[2][1]{\scalebox{#1}{\oldding{#2}}}
\definecolor{applegreen}{rgb}{0.55, 0.71, 0.0}
\definecolor{lavendermagenta}{rgb}{0.93, 0.51, 0.93}

\newenvironment{customlegend}[1][]{%
    \begingroup
    \csname pgfplots@init@cleared@structures\endcsname
    \pgfplotsset{#1}%
}{%
    \csname pgfplots@createlegend\endcsname
    \endgroup
}%
\usetikzlibrary{shapes.geometric}
\pgfdeclareplotmark{mystar}{
    \node[star, star points=5, star point ratio=0.4, draw=black, solid, fill=black, minimum width=3pt, inner sep=0pt, outer sep=0pt, anchor=center] {};
}

\def\addlegendimage{\csname pgfplots@addlegendimage\endcsname}

\usepackage{romannum}
\def\@fnsymbol#1{\ensuremath{\ifcase#1\or *\or \dagger\or \ddagger\or
   \mathsection\or \mathparagraph\or \|\or **\or \dagger\dagger
   \or \ddagger\ddagger \else\@ctrerr\fi}}
\newcommand{\ssymbol}[1]{^{\@fnsymbol{#1}}}

\usepackage{pgfplotstable}
\pgfplotsset{compat=1.9} 

\pgfplotsset{
    legend image with text/.style={
        legend image code/.code={%
            \node[anchor=center] at (0.0cm,0cm) {#1};
        }
    },
}

\definecolor{LimeGreen}{rgb}{0.2, 0.8, 0.2}
\definecolor{ProcessBlue}{rgb}{0.0, 0.72, 0.92}

\newcommand{\cotran}{CoTran\xspace}

\makeatletter
\def\adl@drawiv#1#2#3{%
        \hskip.5\tabcolsep
        \xleaders#3{#2.5\@tempdimb #1{1}#2.5\@tempdimb}%
                #2\z@ plus1fil minus1fil\relax
        \hskip.5\tabcolsep}
\newcommand{\cdashlinelr}[1]{%
  \noalign{\vskip\aboverulesep
           \global\let\@dashdrawstore\adl@draw
           \global\let\adl@draw\adl@drawiv}
  \cdashline{#1}
  \noalign{\global\let\adl@draw\@dashdrawstore
           \vskip\belowrulesep}}
\makeatother

\makeatletter
\patchcmd{\@algocf@start}
  {-1.5em}
  {0pt}
{}{}
\makeatother

\newcommand{\BibTeX}{B\kern-.05em{\sc i\kern-.025em b}\kern-.08em\TeX}

\makeatletter
\newcommand*\mysize{%
  \@setfontsize\mysize{6.95}{9.0}%
}
\newcommand*\mysizeSub{%
  \@setfontsize\mysizeSub{6.0}{9.0}%
}
\newcommand*\mynewsizeSub{%
  \@setfontsize\mynewsizeSub{6.95}{9.0}%
}
\makeatother


\begin{document}


\begin{frontmatter}


\paperid{2573} 


\title{\cotran: An LLM-based Code Translator using Reinforcement Learning with Feedback from \\Compiler and Symbolic Execution}


\author[A]{\fnms{Prithwish}~\snm{Jana}\thanks{Corresponding Author. Email: pjana7@gatech.edu. The paper has been published at the \textit{\nth{27} European Conference on Artificial Intelligence (ECAI-2024)}. This is the full version that includes the supplementary (Appendix).}}
\author[A]{\fnms{Piyush}~\snm{Jha}}
\author[B]{\fnms{Haoyang}~\snm{Ju}}
\author[C]{\fnms{Gautham}~\snm{Kishore}}
\author[D]{\fnms{Aryan}~\snm{Mahajan}}
\author[A]{\fnms{Vijay}~\snm{Ganesh}}

\address[A]{Georgia Institute of Technology, USA}
\address[B]{University of Toronto, Canada}
\address[C]{University of California San Diego, USA}
\address[D]{Columbia University, USA}


\begin{abstract}
    In this paper, we present an LLM-based code translation method and an associated tool called \cotran, that translates whole-programs from one high-level programming language to another. Existing LLM-based code translation methods lack training to ensure that the translated code reliably compiles or bears substantial functional equivalence to the input code. In our work, we fine-tune an LLM using reinforcement learning, incorporating compiler feedback, and symbolic execution (symexec)-based testing feedback to assess functional equivalence between the input and output programs. The idea is to guide an LLM during fine-tuning, via compiler and symexec-based testing feedback, by letting it know how far it is from producing perfect translations. We conduct extensive experiments comparing \cotran with 14 other code translation tools, including human-written transpilers, LLM-based translation tools, and ChatGPT. Using a benchmark of over \num{57000} code pairs in Java and Python, we demonstrate that \cotran outperforms the other tools on relevant metrics such as compilation accuracy (CompAcc) and functional equivalence accuracy (FEqAcc). For example, in Python-to-Java translation, \cotran achieves $48.68\%$ FEqAcc and $76.98\%$ CompAcc, whereas the nearest competing tool (PLBART-base) gets $38.26\%$ and $75.77\%$ respectively. Additionally, \cotran, built on top of CodeT5, improves FEqAcc by $+14.89\%$ and CompAcc by $+8.14\%$ for Python-to-Java (resp., $+12.94\%$ and $+4.30\%$ for Java-to-Python).

\end{abstract}

\end{frontmatter}


\vspace{-1.9mm}
\section{Introduction}
\label{sec1:intro}
Automatic translation of code from one high-level language to another is an important area of software engineering research with applications in code migration and cross-platform interoperability~\cite{grimmer2018cross,mateus2023learning}. Traditional code translation tools, often called \textit{transpilers} rely on human-written rules and, thus, can be quite expensive to develop. 

To address this issue, many researchers~\cite{feng2020codebert,lu2021codexglue} have recently proposed the use of Large Language Models (LLMs)~\cite{vaswani2017attention} for code translation. The rise of LLMs is perhaps the most important development in AI in recent years, and even more remarkably they are being used successfully for many software engineering tasks such as code synthesis~\cite{svyatkovskiy2020intellicode} and code completion~\cite{wang2022compilable}. In this context, LLMs have been used for translating a single function from a source (e.g., Java) to a target language (e.g., Python)~\cite{ahmad2021unified,lu2021codexglue,roziere2020unsupervised}. Unfortunately, function-to-function translation is not sufficient (refer Section~\ref{subsec4:Results}: Finding 8) for whole-program translation tasks. Further, current LLM-based methods lack a proactive approach to ensure that the translated code reliably compiles or bears substantial functional equivalence to the input code. This is a serious problem, hindering the wider adoption of LLM-based code translation techniques.

\begin{figure*}[t!]
  \centering
     \begin{subfigure}{0.495\textwidth}
         \centering
         \includegraphics[height=0.34\textwidth]{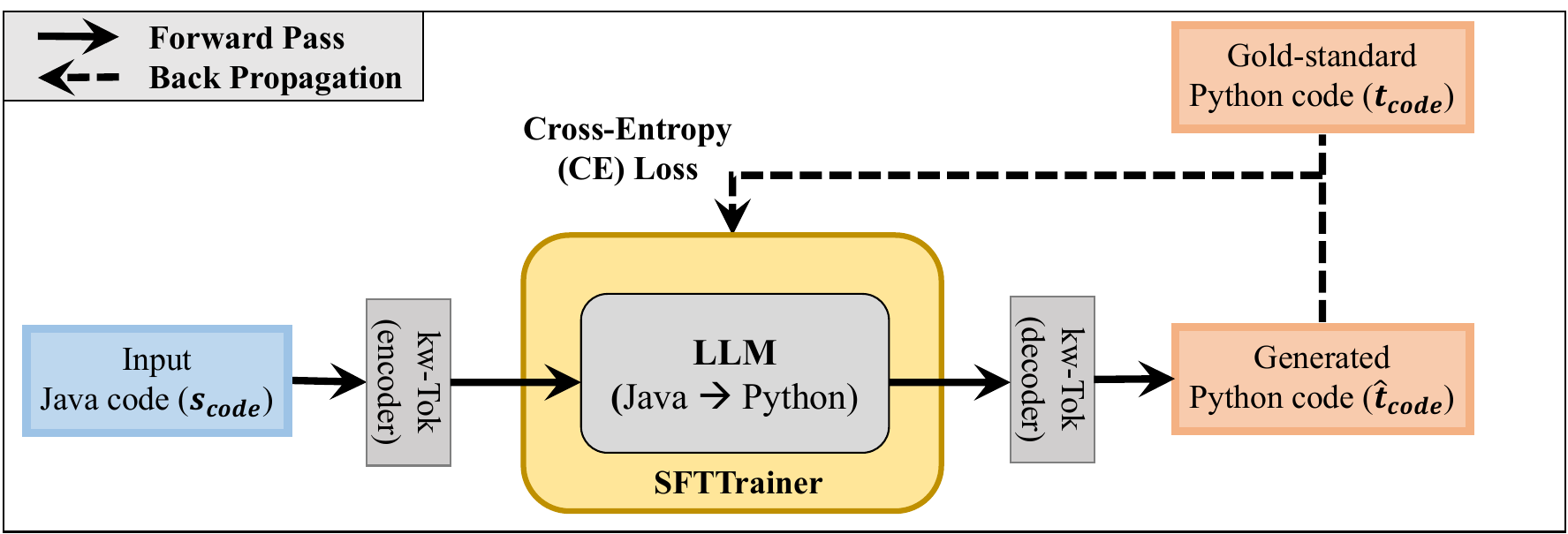}%
         \caption{The baseline \cotran (using Supervised Fine-Tuning i.e., SFT)}
         \label{fig:CoTranBaseline}
     \end{subfigure}
     \begin{subfigure}{0.495\textwidth}
         \centering
         \includegraphics[height=0.339\textwidth]{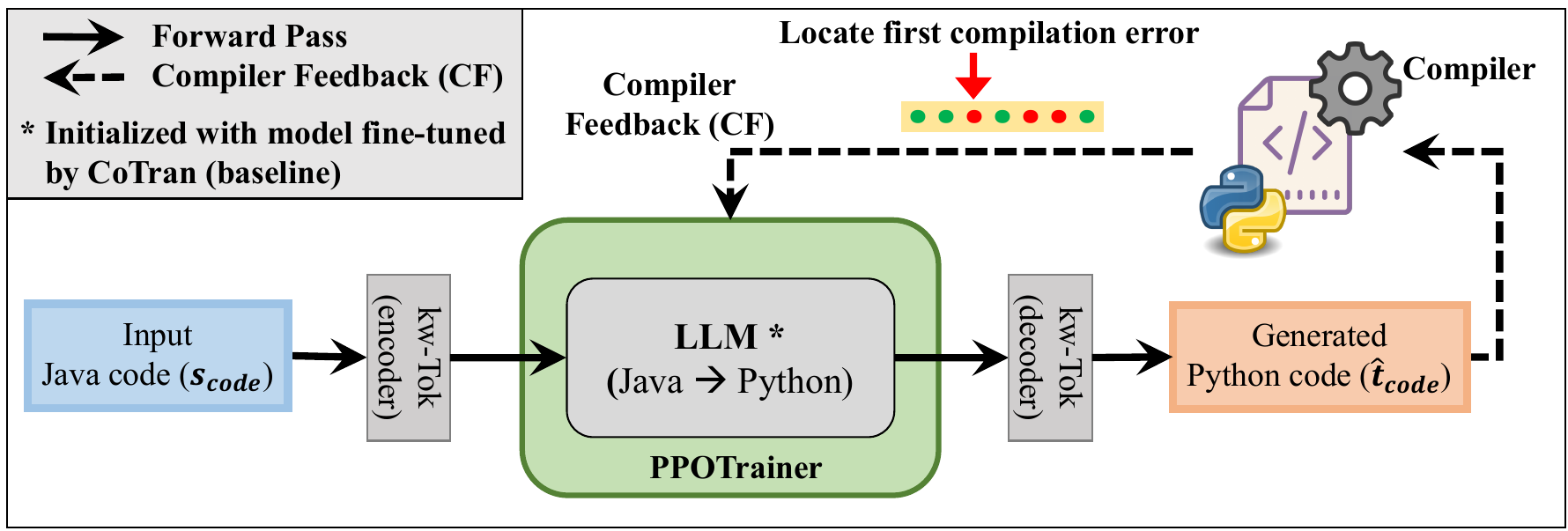}%
         \caption{\cotran + CF (using RL-based fine-tuning by PPO Algorithm)}
         \label{fig:CoTranWithCF}
     \end{subfigure}
     \hfill
     \vspace{3mm}
     \begin{subfigure}[b]{0.71\textwidth}
         \includegraphics[height=0.37\textwidth]{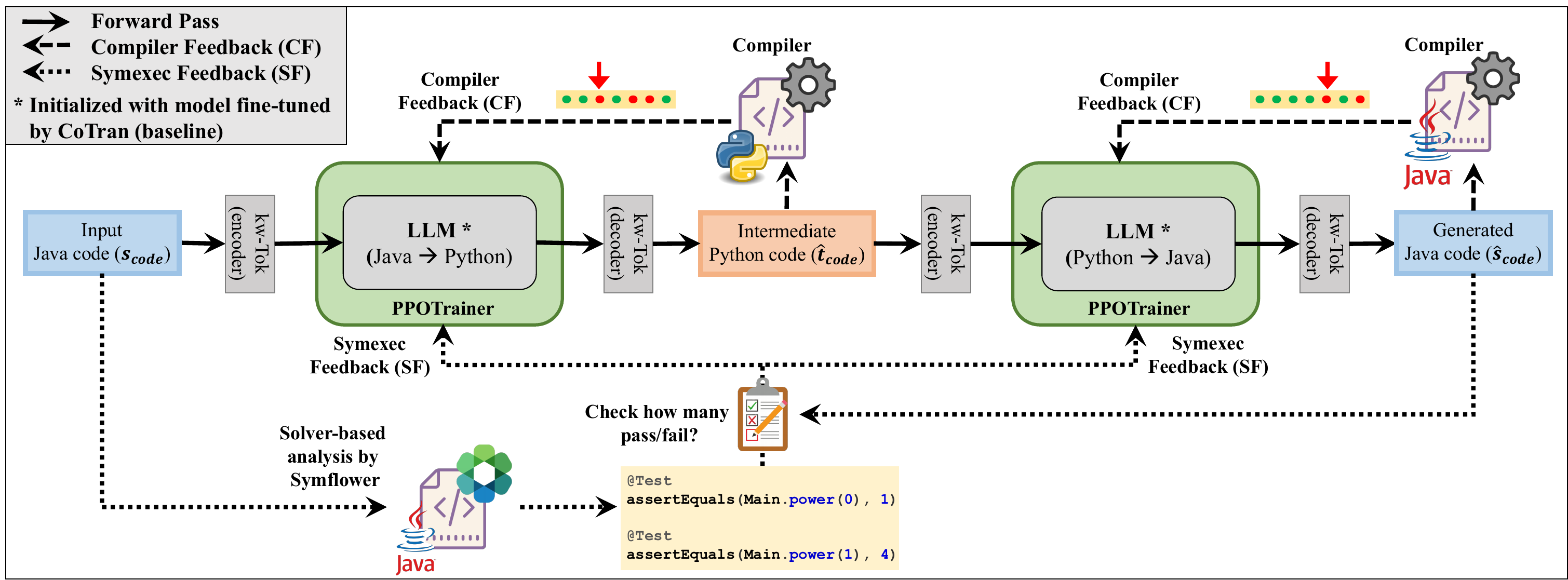}%
         \caption{\cotran + CF + SF (using RL-based fine-tuning on b2b LLMs, refer func. \texttt{\textbf{\textsc{RL\_Bk2Bk}}} in Algorithm~\ref{algo:b2bTrain})}
         \label{fig:CoTranWithCFSF}
     \end{subfigure}
     \begin{subfigure}[b]{0.28\textwidth}
         \includegraphics[height=0.94\textwidth]{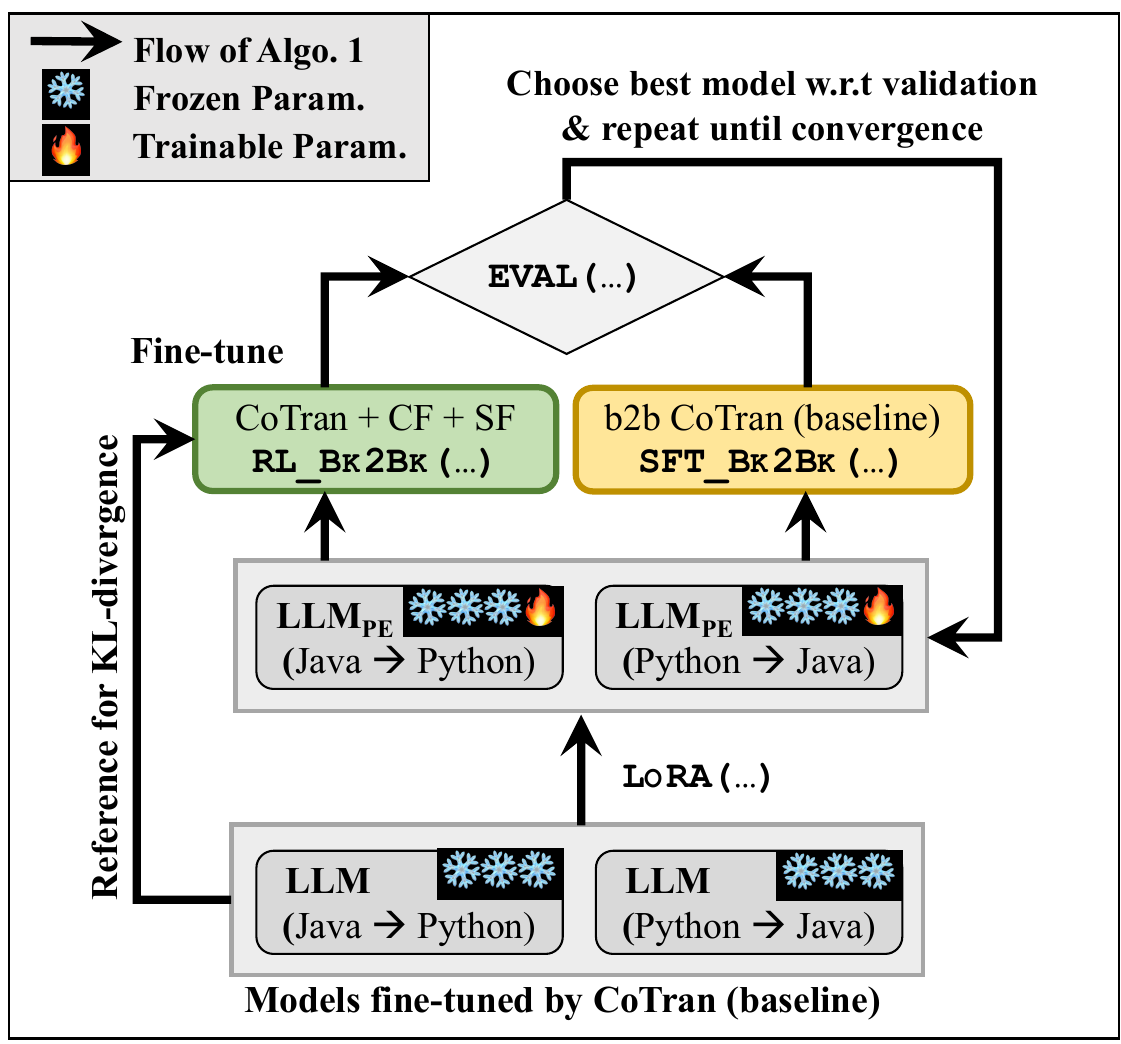}%
         \caption{RL+SFT Interleaved Training Loop}
         \label{fig:CoTranInterleavedLoop}
     \end{subfigure}
     \vspace{4.7mm}
  \caption{{\bf Fine-tuning LLMs with Compiler Feedback (CF) and Symbolic Execution Feedback (SF):} (a) \textbf{\cotran (baseline)} is an LLM fine-tuned without feedback loops. It is fine-tuned to optimize cross-entropy loss, employing the proposed keyword-based tokenizer (kw-Tok). (b) \textbf{\cotran + CF} is an LLM fine-tuned using an RL-based PPOTrainer where the compiler provides feedback in the form of a reward. For P2J, \texttt{javac} compiler is used. For J2P, we use the \texttt{pylint}~\protect\cite{pylint} static code analyzer as Python is an interpreted language. (c) In \textbf{\cotran + CF + SF}, we jointly fine-tune back-to-back (b2b) LLMs (the forward LLM for J2P and the backward LLM for P2J) using RL. Through solver-based analysis of the input Java code, we generate a set of JUnit tests, which are then verified on the output Java code to compute SF. The reward for each LLM is a combination of SF and the respective CF. (d) We begin with the forward (J2P) and backward (P2J) LLMs, fine-tuned by CoTran (baseline) and with frozen parameters. We then continue by fine-tuning a parameter-efficient (PE) version of these LLMs by interleaving RL and SFT, training only a few additional parameters through Low-Rank Adaptation (LoRA)~\protect\cite{hu2021lora} (refer Algorithm~\ref{algo:b2bTrain}).}
  \vspace{5mm}
  \label{fig:compilerAndSolverLoss}
\end{figure*}

We address this problem by modifying the fine-tuning mechanism for code LLMs to incorporate feedback from compiler and symbolic execution (symexec)-based testing. Our method retains the positive aspect of LLM-based code translation, while improving its accuracy dramatically. The core idea behind our method is to employ \textit{corrective feedback loops}~\cite{ganesh2022machine, jana2024neurosymbolic} between learning (the LLM) and reasoning (compiler \& symexec), for which we have a two-fold strategy. Firstly, if the output code produced by the LLM does not compile, a \textbf{compiler feedback (CF)} is computed. Secondly, if the output and input codes are not equivalent (w.r.t. a symexec-generated test suite), a suitable \textbf{symexec feedback (SF)} is computed. This feedback is then used in a reinforcement learning (RL) framework to fine-tune the LLM. Our results on \textbf{Java-to-Python (J2P)} and \textbf{Python-to-Java (P2J)} translation indicate that these new feedback functions significantly improve accuracy compared to previous methods. Another interesting feature is that we use two back-to-back LLMs, one for source-to-target ($S\to T$) and another for target-to-source ($T\to S$), that are fine-tuned together. This setup of complementary models simplifies the automation of functional equivalence checking between the input (of $S\to T$ model) and output (of $T\to S$ model), both in $S$.

\vspace{0.9mm}
\noindent{\bf Contributions.}
\vspace{-3mm}
\begin{itemize}[leftmargin=0.4cm]
\setlength{\parskip}{0pt}
\setlength\itemsep{0.2em}
    \item We describe a whole-program translation method (and tool \cotran) based on interleaved training of LLMs through Supervised Fine-Tuning (SFT)- and RL-based optimization, to incorporate CF and SF into the fine-tuning process. Currently, \cotran is configured to translate between Java and Python (J2P and P2J). It can be effortlessly adapted for other language pairs ($S$, $T$) given: \emph{(i)} a dataset of equivalent code pairs in $S$ and $T$, \emph{(ii)} compilers for both languages and \emph{(iii)} an automatic test-case generation tool for either $S$ or $T$ -- all of which are readily available for popular languages. A key insight is that LLMs trained for software engineering tasks (e.g., code synthesis) can greatly benefit from feedback via compilers, program analysis, and test generation tools. (Section~\ref{subsec3:training})
    \item Our fine-grained compiler feedback (CF) guides an LLM during fine-tuning, helping it assess the proximity of generated translations to a perfectly compiling one. It is much more effective than a Boolean yes/no feedback e.g., for P2J, \cotran achieves +11.57\% higher functional equivalence accuracy and +17.36\% higher compilation accuracy compared to PPOCoder~\cite{shojaee2023execution}. (Section~\ref{subsec3:loss})
    \item Our symexec feedback (SF) is used to fine-tune $S \to T \to S$ back-to-back LLMs. We generate unit tests on the input code and check them on the output translated code, to provide feedback on their \textit{inequivalence}. Our method is entirely agnostic to any specific test-case generation tool. As far as we know, no existing methods utilize symexec-based test generation for functional equivalence checking and associated feedback function. (Section~\ref{subsec3:loss})
     \item We perform an extensive empirical evaluation and ablation study of \cotran against 11 state-of-the-art LLM-based translation tools and 3 human-written transpilers. A range of metrics are assessed incl., functional equivalence\footnote{Note that program equivalence is an undecidable problem in general, and we make no claims about establishing program equivalence. By functional equivalence of two codes, we mean something more limited, namely, that they produce equivalent results w.r.t. a test suite.} accuracy (FEqAcc), compilation accuracy (CompAcc), BLEU~\cite{papineni2002bleu}, CodeBLEU~\cite{ren2020codebleu} and the proposed average first error position ($\text{errPos}_{\nth{1}}$). \cotran and its variants outperform all other tools of similar size for both J2P and P2J. Compared to the nearest competing tool on FEqAcc, CoTran gets +9.62\% (vs. PPOCoder~\cite{shojaee2023execution}) in J2P and +10.42\% (vs. PLBART-base~\cite{ahmad2021unified}) in P2J. \cotran even outperforms ChatGPT~\cite{chatgpt} (a much larger model) on all the metrics in P2J and all but one metric (FEqAcc) in J2P.  
     Furthermore, our tool outperforms transpilers, e.g., for J2P, \cotran gets +{53.43}\% in FEqAcc, +{38.84}\% in CompAcc, and +{38.47}\% in $\text{errPos}_{\nth{1}}$ (vs. TSS CodeConv~\cite{tangiblesoftsoln}). (Section~\ref{sec4:experiments})

    \item We introduce a specialized Keyword Tokenizer (kw-Tok) for code translation, boosting FEqAcc by $+3.57\%$ for J2P and $+6.62\%$ for P2J, and CompAcc by $+3.28\%$ and $+4.79\%$, resp. (Section~\ref{sec3:methodology})

    \item We contribute a large, well-curated dataset AVATAR-TC (built on top of AVATAR~\cite{ahmad-2021-avatar}) having \num{57000}+ Java-Python code pairs with human-written test-cases (TCs). Compared to AVATAR, faulty codes are manually fixed and TCs (by problem-setters) are collected from coding platforms, enabling the calculation of translator FEqAcc. Based on what we know, this is the first large-scale dataset thoroughly testing code pairs with human-written TCs.
    (Section~\ref{sec4:avatartc})
\end{itemize}

\section{Related Work}
\label{sec2:rel-work}
\noindent{\bf Rule-based Transpilers.} Rule-based handcrafted transpilers are usually built using traditional compiler techniques and concepts such as parsing and abstract syntax trees. Examples of such transpilers include \texttt{java2python}~\cite{java2python} and \texttt{py2java}~\cite{py2java}. TSS code converter~\cite{tangiblesoftsoln} is a commercial J2P transpiler. Overall, these transpilers vary by the intricacies and difficulty level of constructs (e.g., lambda, anonymous inner class). Many such tools state a disclaimer with limitations that the translated codes should not be expected to compile and run readily.


\noindent{\bf Transformer and ML-based Code Translation Tools.} The encoder-decoder architecture of Transformers revolutionized Natural Language (NL) translation, using contextualized representations of words~\cite{vaswani2017attention}. This led to the  evolution of advanced language models for code translation task, such as encoder-only CodeBERT~\cite{feng2020codebert} and decoder-only CodeGPT~\cite{lu2021codexglue}. CodeT5~\cite{wang2021codet5} is an encoder-decoder architecture that leverages code semantics from developer-assigned identifiers. Another tool is PLBART~\cite{ahmad2021unified} which is a unified transformer trained through denoising autoencoding and performs a range of tasks among NL and Programming Language (PL). Conversely, tree-to-tree~\cite{chen2018tree} is an attention-based LSTM neural network.

\noindent{\bf LLM-based methods using compiler or unit testing.} TransCoder-ST~\cite{roziere2022leveraging} is an unsupervised code translation tool that uses self-training and automated unit tests to verify source-target code equivalence. However, it uses unit tests to create a synthetic dataset of equivalent codes in two languages, instead of incorporating it for model improvement during training. Recently, Boolean compiler feedback was used to fine-tune LLM for code generation in an RL-based scheme~\cite{wang2022compilable}. PPOCoder~\cite{shojaee2023execution} is an RL-based code translation framework that additionally uses CodeBLEU-inspired feedback in the LLM fine-tuning process. While both the tools use the same Boolean compiler feedback that is true if the generated code compiles and false otherwise, our proposed compiler feedback function is fine-grained -- it captures how far the translation is from a perfectly compilable code. Back-to-back (b2b) code translation although proposed earlier~\cite{roziere2020unsupervised}, is not leveraged for symexec-based functional equivalence checking. 

\section{The \cotran Method}
\label{sec3:methodology}
\noindent{\bf The Code Translation Learning Problem.} Let $S$ denote the source language and $T$ the target language. The code translation learning problem is to learn a language translator $f_{ST}:S\to T$, such that the output of the translator is a $T$-program that is \textit{syntactically correct} (as per the grammar of $T$) and is {\it functionally equivalent} to the input $S$-program (input-output equivalence w.r.t. a test suite). 

\noindent{\bf Brief Overview of our Method.}\label{subsec3:overview} Please refer to Figure~\ref{fig:compilerAndSolverLoss} for an architectural overview of the \cotran tool. In order to learn the language translator, we build off a sequence-to-sequence (seq2seq) attention-based transformer model consisting of an encoder and decoder stack~\cite{vaswani2017attention}. For the remainder of the paper, we use the term Large Language Models (LLMs) to refer to such models. The proposed fine-tuning pipeline consists of two steps. In the first step, we separately perform supervised fine-tuning (SFT) on $\text{LLM}_{f}:S\to T$ and $\text{LLM}_{b}:T\to S$ (the forward and backward models respectively) to optimize \textit{cross-entropy loss}. In the second step, we jointly fine-tune these two models in a parameter-efficient way on the back-to-back translation task i.e., $S\to T\to S$. For this process, use an RL+SFT interleaved training loop that incorporates \textit{compiler-} and \textit{symexec-feedback}. Figure~\ref{fig:compilerAndSolverLoss} illustrates the feedback mechanisms.

\noindent{\bf Keyword Tokenizer (kw-Tok).}
\label{subsec3:preprocessing}
The code $\mathbf{s}_{code} \in S$ is first tokenized based on the syntax (grammar) of the corresponding language. Next, the code-specific tokens thus obtained, are converted to a sequence of token-ids using a RoBERTa tokenizer, which is modified for the PL translation task. To avoid splitting the PL keywords into multiple subtokens, we add all the keywords (e.g., \texttt{volatile}, \texttt{transitive} in Java; \texttt{elif}, \texttt{instanceof} in Python) and operators (e.g., \texttt{>=}, \texttt{**=}) of $S$ and $T$ to the tokenizer vocabulary. These are collected from the list of terminal symbols of these languages' formal grammar. Grammars-v4~\cite{parrGrammarsv4} provides a collection of formal grammars for most common languages in use.
We also add special tokens to the vocabulary e.g., \texttt{NEW\_LINE}, \texttt{INDENT}, \texttt{DEDENT} for Python. We refer to this modified RoBERTa tokenizer as \textit{Keyword Tokenizer (kw-Tok)}. This modification ensures that \textit{kw-Tok} generates a \textit{single token-id} for all the special tokens, keywords, and operators of languages $S$ and $T$. Thus, given a training example $(\mathbf{s}_{code}, \mathbf{t}_{code})\in$ dataset $D$, the tokenizer preprocesses these to a sequence of token-ids viz., $\mathbf{s}=(s_1, s_2, \ldots, s_n)$ and $\mathbf{t}=(t_1, t_2, \ldots, t_m)$. The LLM ($S\to T$) takes the source sequence $\mathbf{s}$ and generates a target sequence $\widehat{\mathbf{t}}=(\widehat{t}_1, \widehat{t}_2, \ldots, \widehat{t}_q)$.

\subsection{Definitions: CE Loss, CF, and SF}
\label{subsec3:loss}

Refer to Figure~\ref{fig:compilerAndSolverLoss} for a pictorial representation of cross-entropy (CE) loss, compiler feedback (CF), and symexec feedback (SF). Let $\theta_f$ and $\theta_b$ be the trainable parameters for $\text{LLM}_{f}$ and $\text{LLM}_{b}$ respectively.

\noindent{\bf Cross-Entropy (CE) Loss.} As is typical~\cite{wang2021codet5,vaswani2017attention} in translation tasks, LLMs are fine-tuned to minimize the cross-entropy (CE) loss through supervised learning. For the forward LLM that generates $\widehat{\mathbf{t}}$ given a tokenized training instance $(\mathbf{s}, \mathbf{t})$, the CE loss is defined as:

\begin{equation}
\mathcal{L}^{\theta_f}_{CE}\left(\mathbf{t}, \widehat{\mathbf{t}}\,\right)=-\frac{1}{\ell}\sum_{i=1}^{\ell}\sum_{j=1}^{\left|V\right|}\mathbb{1}_{ij}\log P^{\theta_f}_{ij}
\label{eq:CEloss}
\end{equation}

\noindent where, $\ell=\max\left(\left|\,\widehat{\mathbf{t}}\,\right|, \left|\mathbf{t}\right|\right)$ and $V$ is the tokenizer vocabulary. $\mathbb{1}_{ij}$ is $1$ \textit{iff} the $i^{\text{th}}$ token in reference translation $\mathbf{t}$ is the $j^{\text{th}}$ word of $V$ and, $P^{\theta_f}_{ij}$ is the probability that the $i^{\text{th}}$ token in predicted translation $\widehat{\mathbf{t}}$ is the $j^{\text{th}}$ word of $V$. The CE loss is, however, more suitable for machine translation of NLs than PLs. For NLs, the grammar is lenient, and an approximate translation is oftentimes good enough. But in PL translation where the grammar is strict and functional equivalence is paramount, the aim is to generate compilable codes in the target language and, to make sure that $\widehat{\mathbf{t}}$ is functionally equivalent to $\mathbf{s}$.

\noindent{\bf Compiler Feedback (CF).} We use a compiler to assess the syntactic correctness of an LLM-generated translation during fine-tuning and in turn, provide feedback to the LLM. This guides the model in determining how the translation fares relative to a perfectly compilable code. For CF, the position of the first compilation error serves as a key heuristic in evaluating syntactic correctness. If it appears near the end of the code, it typically indicates fewer cascading errors and a closer-to-perfect compilation. So, for the forward LLM that generates $\widehat{\mathbf{t}}$ given a tokenized instance $(\mathbf{s}, \mathbf{t})$, we formulate the feedback as:

\begin{equation}
\omega_{\text{compiler}}\left(\;\widehat{\mathbf{t}}\;\right) = \begin{cases} 
      +2 &  ,\text{if }\widehat{\mathbf{t}}\text{ compiles} \\
      \frac{f\left(\text{compiler}_T, \widehat{\mathbf{t}}\,\right)}{\left|\,\widehat{\mathbf{t}}\,\right|+1} & ,\text{otherwise}
\end{cases}
\label{eq:compilerOnlyFeedback}
\end{equation}

\noindent where, the function $f(\cdot, \cdot)$ applies the $T$-language compiler on $\widehat{\mathbf{t}}$ and returns the token position $\left(\in \left[1, |\,\widehat{\mathbf{t}}\,|\right]\right)$ of the first syntax error. So, $\omega_{\text{compiler}}$ is $+2$ for a perfect compilation and goes on decreasing from $+1$ to $0$ as the token position of the first syntax error is closer to the beginning of the code. However, the LLM should not game the system to generate $\widehat{\mathbf{t}}$ as a dummy code with no compilation error e.g., a Hello-World program. To penalize such cases we posit that $\widehat{\mathbf{t}}$ should also be close-by in length to $\mathbf{t}$. Accordingly, CF is defined as the product of $\omega_{\text{compiler}}$ and the value of a $(0,1]$ Gaussian distribution at $|\,\widehat{\mathbf{t}}\,|$, which is centered at $|\mathbf{t}|$ and has a standard deviation of $\frac{\left|\mathbf{t}\right|}{4}$, as follows:

\begin{equation}
\omega_{CF}\left(\mathbf{t}, \widehat{\mathbf{t}}\,\right) = \omega_{\text{compiler}}\left(\,\widehat{\mathbf{t}}\,\right) \times e^{-\frac{1}{2}\left(\frac{\left|\,\widehat{\mathbf{t}}\,\right|-\left|\mathbf{t}\right|}{\left|\mathbf{t}\right|/4}\right)^2}
\label{eq:CF}
\end{equation}

\noindent{\bf Symbolic Execution Feedback (SF).} We also introduce a \textit{symexec feedback} (SF) that lets the LLM during fine-tuning know the extent to which the generated translation is \textit{(in)-equivalent}\,\footnote{It goes without saying that one cannot guarantee functional equivalence between two programs with a purely testing approach. However, we can establish inequivalence via a sufficient amount of testing.} to the source-language code. As it can be challenging to assess functional inequivalence among two codes of different languages ($\mathbf{s}$ and $\widehat{\mathbf{t}}$), we use two back-to-back (b2b) LLMs: $S\to T$ and $T\to S$ such that the same test suite can be used for inequivalence testing. Given a tokenized training instance $(\mathbf{s}, \mathbf{t})$, the forward LLM ($S\to T$) translates $\mathbf{s}$ to an intermediate target-language sequence $\widehat{\mathbf{t}}$, which the backward LLM ($T\to S$) translates back to $\widehat{\mathbf{s}}$ in an attempt to reconstruct $\mathbf{s}$. To compute SF, we use solver-based analysis by an industrial-strength symexec tool called Symflower~\cite{symflowerbib}. Through symexec, Symflower generates essential unit tests even for functions involving complex data types. It computes the necessary inputs and expected outputs to cover all linearly independent control-flow paths within a function. For J2P, $S$ is Java and $T$ is Python. We use Symflower to automatically generate JUnit tests ($\mathcal{J}_{\mathbf{s}}$) for each function in $\mathbf{s}\in S$. These unit tests are checked on the corresponding function in $\widehat{\mathbf{s}}$ and both LLMs get feedback on how many tests pass. Accordingly, SF is defined as: 

\begin{equation}
\omega_{SF}\left(\mathbf{s}, \widehat{\mathbf{s}}\right) = \frac{\epsilon + \sum\limits_{j\in \mathcal{J}_{\mathbf{s}}}\mathbb{1}_{j(\widehat{\mathbf{s}})\equiv\text{Success}}}{\epsilon + \left|\mathcal{J}_{\mathbf{s}}\right|}
\label{eq:SF}
\end{equation}

\noindent where, $\epsilon$ is a small positive value and $\mathbb{1}_{j(\widehat{\mathbf{s}})\equiv\text{Success}}$ is $1$ when the $j^{\text{th}}$ JUnit test on $\mathbf{s}$ successfully passes on $\widehat{\mathbf{s}}$, else it is $0$. Our interleaved training loop (Section~\ref{subsec3:training}) incorporates CF on individual LLMs, in addition to SF. So, a feedback on $(\mathbf{s}$, $\widehat{\mathbf{s}})$-inequivalence with a feedback on compilability of $\widehat{\mathbf{t}}$ and $\widehat{\mathbf{s}}$, correlates to $(\mathbf{s}$, $\widehat{\mathbf{t}}\,)$-inequivalence.

\subsection{The \cotran Training Loop}
\label{subsec3:training}

Given a training instance $(\mathbf{s}_{code}, \mathbf{t}_{code})\in$ dataset $D$, kw-Tok converts them into token-ids, $\mathbf{s}=(s_1, s_2, \ldots, s_n)$ and $\mathbf{t}=(t_1, t_2, \ldots, t_m)$. We outline a two-step process for fine-tuning the LLMs.

\noindent{\bf Supervised fine-tuning of translation models, $\mathbf{\text{LLM}_{f}}$ and $\mathbf{\text{LLM}_{b}}$:} In the first phase, we separately perform SFT on $\text{LLM}_{f}$ and $\text{LLM}_{b}$ to minimize the CE loss over the respective translation tasks. $\text{LLM}_{f}$ translates $\mathbf{s}$ to a $T$-sequence $\widehat{\mathbf{t}}=(\widehat{t}_1, \widehat{t}_2, \ldots, \widehat{t}_q)$. Through SFT, it is optimized to minimize the CE loss between $\widehat{\mathbf{t}}$ and $\mathbf{t}$ i.e., $\mathcal{L}^{\theta_f}_{CE}(\mathbf{t}, \widehat{\mathbf{t}}\,)$ as defined in Eqn.~\ref{eq:CEloss}. Similarly, $\text{LLM}_{b}$ translates $\mathbf{t}$ to a $S$-sequence $\widehat{\mathbf{s}}=(\widehat{s}_1, \widehat{s}_2, \ldots, \widehat{s}_r)$ and is optimized to minimize $\mathcal{L}^{\theta_b}_{CE}\left(\mathbf{s}, \widehat{\mathbf{s}}\right)$.

\noindent{\bf Jointly fine-tuning $\mathbf{\text{LLM}_{f}}$ and $\mathbf{\text{LLM}_{b}}$ using interleaved RL+SFT:} Next, we further fine-tune $\text{LLM}_{f}$ and $\text{LLM}_{b}$, but this time together through back-to-back (b2b) translation in order to incorporate CF and SF. Given a source-language code, a successful code translation calls for generating a target-language code that \textit{compiles} and is \textit{functionally equivalent} to the input code. Here, optimizing LLMs with CE is not sufficient. So, we defined feedback mechanisms (CF and SF) to inform an LLM during fine-tuning about its proximity to achieving a perfect translation. However as CF and SF are non-differentiable functions, they cannot be directly used as loss functions to fine-tune an LLM. So, it is essential to construct an RL setting. 

\SetInd{0.3em}{0.3em}
\SetAlFnt{\footnotesize}
\SetKwRepeat{Do}{do}{while}
\SetKwInOut{KwIn}{Input}
\SetKwInOut{KwOut}{Output}
\newcommand\mycommfont[1]{\scriptsize\ttfamily\textcolor{blue}{#1}}
\setlength{\algomargin}{1em}
\begin{algorithm}[!t]
\SetNoFillComment
\SetCommentSty{mycommfont}
\setstretch{0.92}
    \caption{RL+SFT Interleaved Training for \cotran}
    \label{algo:b2bTrain}
    \KwIn{$\text{M}_{f}$ (forward LLM); $\text{M}_{b}$ (backward LLM); $Tok$ (kw-Tok); \texttt{tP($\cdot$)} (trainable param); $D_{\text{trn}}, D_{\text{val}}$ (training \& validation data)}
    \KwOut{Learned LLMs $\text{M}_{f}$ (for $S\to T$) and $\text{M}_{b}$ (for $T\to S$)}
    \algrule
    \SetKwFunction{RL}{\textbf{\textsc{RL\_Bk2Bk}}}
    \SetKwProg{Fn}{Function}{:}{}
    \Fn{\RL{$\text{LLM}_{\text{f}}$, $\text{LLM}^{\text{ref}}_{\text{f}}$, $\text{LLM}_{\text{b}}$, $\text{LLM}^{\text{ref}}_{\text{b}}$, PPO$_{\text{f}}$, PPO$_{\text{b}}$, $D$}}{ 
        \For{$\text{epoch}\in[1,E]$}{\ForEach{$(\mathbf{s}, \mathbf{t})\in D$}{
        $\widehat{\mathbf{t}} \gets \text{LLM}_{\text{f}}(\mathbf{s})$;\,\,$\widehat{\mathbf{s}} \gets \text{LLM}_{\text{b}}(\,\widehat{\mathbf{t}}\,)$\\
            $r_{\text{f}} \gets \omega_{\text{CF}}(\mathbf{t}, \widehat{\mathbf{t}}\,) + \omega_{\text{SF}}\left(\mathbf{s}, \widehat{\mathbf{s}}\right) - \beta \cdot d_{\text{KL}}(\text{LLM}_{\text{f}}, \text{LLM}^{\text{ref}}_{\text{f}})$\\
            $r_{\text{b}} \gets \omega_{\text{CF}}\left(\mathbf{s}, \widehat{\mathbf{s}}\right) + \omega_{\text{SF}}\left(\mathbf{s}, \widehat{\mathbf{s}}\right) - \beta \cdot d_{\text{KL}}(\text{LLM}_{\text{b}}, \text{LLM}^{\text{ref}}_{\text{b}})$\\
            BackProp: $\text{LLM}_{\text{f}} \gets \underset{\theta_{\text{f}}}{\text{PPO}_{\text{f}}}\left(\text{LLM}_{\text{f}}, r_{\text{f}}\right)$ \tcp*{$\theta_{\text{f}}$: tP($\text{LLM}_{\text{f}}$)}
            BackProp: $\text{LLM}_{\text{b}} \gets \underset{\theta_{\text{b}}}{\text{PPO}_{\text{b}}}\left(\text{LLM}_{\text{b}}, r_{\text{b}}\right)$ \tcp*{$\theta_{\text{b}}$: tP($\text{LLM}_{\text{b}}$)}
        }}
        \Return $\text{LLM}_{\text{f}}$, $\text{LLM}_{\text{b}}$
    }
    \SetKwFunction{SFT}{\textbf{\textsc{SFT\_Bk2Bk}}}
    \SetKwProg{Fn}{Function}{:}{}
    \Fn{\SFT{$\text{LLM}_{\text{f}}$, $\text{LLM}_{\text{b}}$, Adam$_{\text{f}}$, Adam$_{\text{b}}$, D}}{
        \For{$\text{epoch}\in[1,E]$}{ \ForEach{$(\mathbf{s}, \mathbf{t})\in D$}{
            $\widehat{\mathbf{t}} \gets \text{LLM}_{\text{f}}(\mathbf{s})$; \,\,$\widehat{\mathbf{s}} \gets \text{LLM}_{\text{b}}(\,\widehat{\mathbf{t}}\,)$\\
            BackP: $\text{LLM}_{\text{f}} \gets \underset{\theta_{\text{f}}}{\text{Adam}_{\text{f}}}\left(\text{LLM}_{\text{f}}, \mathcal{L}^{\theta_{\text{f}}}_{CE}\left(\mathbf{t}, \widehat{\mathbf{t}}\right)\right)$\tcp*{$\theta_{\text{f}}$: tP($\text{LLM}_{\text{f}}$)}
            BackP: $\text{LLM}_{\text{b}} \gets \underset{\theta_{\text{b}}}{\text{Adam}_{\text{b}}}\left(\text{LLM}_{\text{b}}, \mathcal{L}^{\theta_{\text{b}}}_{CE}\left(\mathbf{s}, \widehat{\mathbf{s}}\right)\right)$\tcp*{$\theta_{\text{b}}$: tP($\text{LLM}_{\text{b}}$)}
        }}
        \Return $\text{LLM}_{\text{f}}$, $\text{LLM}_{\text{b}}$
    }
    \SetKwFunction{LoRA}{\textbf{\textsc{LoRA}}}
    \SetKwProg{Fn}{Function}{:}{}
    \Fn{\LoRA{LLM}}{
        $\text{LLM}^{'}$ $\gets$ deep copy of $\text{LLM}$ \tcp*{$\theta$: tP($\text{LLM}$), $\theta'$: tP($\text{LLM}'$)}
        \ForEach{query/value layer $L_i \in \mathbb{R}^{d_{\text{in}}\times d_{\text{out}}}$ of $\text{LLM}$}
        {Initialize projection-up layer $A_i\in \mathbb{R}^{d_{\text{in}}\times r}$ randomly from $\mathcal{N}(0,1)$, projection-down layer $B_i \gets \{0\}^{r\times d_{\text{out}}}$\\
        Replace $L_i$ of $\text{LLM}^{'}$ with $L_i + \frac{\alpha}{r}(A_i\times B_i)$}
        $\theta^{'} \gets \theta \cup A \cup B$. Freeze params in $\theta^{'}$ except $A, B$\\
        \Return $\text{LLM}^{'}$
    }
    \SetKwFunction{eval}{\textbf{\textsc{EVAL}}}
    \SetKwProg{Fn}{Function}{:}{}
    \Fn{\eval{$\text{LLM}_{\text{f}}$, $\text{LLM}_{\text{b}}$}}{
        \Return $\omega_{\text{SF}}(\mathbf{s}, \widehat{\mathbf{s}})$ averaged over all $\mathbf{s}$ in the tokenized $D_{\text{val}}$, \\\phantom{aaaaaaa}with $\text{LLM}_{\text{f}}$, $\text{LLM}_{\text{b}}$ as back-to-back models
    }

    \vspace{1mm}
    \nonl 
    ---------------------------------\textsc{Main Function}---------------------------------

    \vspace{1mm}
    $\text{M}^{'}_{f}$, $\text{M}^{'}_{b}$ $\gets$ \LoRA{$\text{M}_{f}$}, \LoRA{$\text{M}_{b}$}\\
    Initialize RL Optimizers: $\text{PPO}_f, \text{PPO}_b$ for $\text{M}^{'}_{f}, \text{M}^{'}_{b}$\\
    Initialize SFT Optimizers: $\text{Adam}_f, \text{Adam}_b$ for $\text{M}^{'}_{f}, \text{M}^{'}_{b}$\\
    $D^{Tok}_{\text{trn}} \gets \left[\left(Tok(\mathbf{s}_{code}), Tok(\mathbf{t}_{code})\right) \text{for} \left(\mathbf{s}_{code}, \mathbf{t}_{code}\right) \text{in} \;D_{\text{trn}} \right]$\\
    \Do{$\text{valAcc}_{\text{RL}}$, $\text{valAcc}_{\text{SFT}}$ converged or maxEpochs reached}
    {$\text{M}^{\text{SFT}}_{f}$, $\text{M}^{\text{SFT}}_{b}$ $\gets$ \SFT{$\text{M}^{'}_{f}$, $\text{M}^{'}_{b}$, Adam$_{f}$, Adam$_{b}$, $D^{Tok}_{\text{trn}}$}\\
    $\text{M}^{\text{RL}}_{f}$, $\text{M}^{\text{RL}}_{b}$ $\gets$ \RL{$\text{M}^{'}_{f}$, $\text{M}_{f}$, $\text{M}^{'}_{b}$, $\text{M}_{b}$, PPO$_{f}$, PPO$_{b}$, $D^{Tok}_{\text{trn}}$}\\
    $\text{valAcc}_{\text{\text{SFT}}}$, $\text{valAcc}_{\text{RL}}$ $\gets$ \eval{$\text{M}^{\text{SFT}}_{f}$, $\text{M}^{\text{SFT}}_{b}$}, \eval{$\text{M}^{RL}_{f}$, $\text{M}^{RL}_{b}$}\\
    $\text{M}^{'}_{f}$, $\text{M}^{'}_{b}$ $\gets$ $\left(\text{valAcc}_{\text{RL}}\geq \text{valAcc}_{\text{SFT}}\right)$ \textbf{\texttt{?}} $\text{M}^{\text{RL}}_{f}$, $\text{M}^{\text{RL}}_{b}$ \textbf{\texttt{:}} $\text{M}^{\text{SFT}}_{f}$, $\text{M}^{\text{SFT}}_{b}$\\
    }
\end{algorithm}

The scheme for jointly fine-tuning $\text{LLM}_{f}$ and $\text{LLM}_{b}$ is given in Algorithm~\ref{algo:b2bTrain}. In short, we interleave RL-based fine-tuning by Proximal Policy Optimization (PPO)~\cite{schulman2017proximal} and SFT-based optimization of CE loss by Adam optimizer. First, $\text{LLM}_{f}$ ($S\to T$) translates $\mathbf{s}$ to generate an intermediate $T$-sequence $\widehat{\mathbf{t}}$. With this as input, $\text{LLM}_{b}$ ($T\to S$) tries to reconstruct sequence $\mathbf{s}$ and generates a $S$-sequence $\widehat{\mathbf{s}}$. In RL-based fine-tuning, the reward for $\text{LLM}_{f}$ and $\text{LLM}_{b}$ is the sum of SF computed among  $\mathbf{s}$ and $\widehat{\mathbf{s}}$ and the respective CF among $\mathbf{t}$, $\widehat{\mathbf{t}}$ and $\mathbf{s}$, $\widehat{\mathbf{s}}$. To ensure that the LLMs being fine-tuned do not diverge much from the reference LLMs we started with, a KL-divergence~\cite{palenicek2021survey} ($d_{\text{KL}}$) term is subtracted from the reward. This ensures that the PPO algorithm does not over-optimize and is appropriately penalized when the trained model starts to diverge too much from their references. Conversely, for the SFT-based optimization, the forward and backward models are fine-tuned by back-propagating the respective CE losses between the predicted translation and the corresponding gold-standard translation.

Further, as jointly fine-tuning both LLMs is resource-intensive we follow a parameter-efficient approach. For every linear layer of these LLMs that is either query or value, we create Low-Rank Adaptation~\cite{hu2021lora} layers viz., a projection-up matrix $A$ and projection-down matrix $B$, whose matrix-multiplication is initially zero. Except those in $A$ and $B$, all the original parameters of $\text{LLM}_{f}$ and $\text{LLM}_{b}$ are frozen -- making the fine-tuning process space-time efficient.


\begin{table}[t]
  \caption{\textbf{Benchmark Suite:} Statistics of the AVATAR-TC dataset}
 \label{table:avatarModeifiedStats}
  \centering
  \begin{adjustbox}{width=\columnwidth}
  \begin{tabular}{@{}L{2cm}R{0.9cm}R{0.7cm}R{0.7cm}R{0.9cm}R{0.7cm}R{0.7cm}@{}}
    \toprule
    \multirow{2}{*}{Sub-dataset} &
      \multicolumn{3}{c}{\# problem-stmts with test-cases} &
      \multicolumn{3}{c}{\# pairs of  Java-Python codes}\\
      \cmidrule(lr){2-4}\cmidrule(lr){5-7}
      & Train & Valid & Test & Train & Valid & Test\\
    \midrule
    Aizu & \num{762} & \num{41} & \num{190} & \num{14019} & \num{41} & \num{190} \\
    AtCoder     & \num{619} & \num{19} & \num{97} & \num{13558} & \num{19} & \num{97}  \\
    Codeforces     & \num{1625} & \num{96} & \num{401} & \num{23311} & \num{96} & \num{401} \\
    Google CodeJam     & \num{59} & \num{1} & \num{4} & \num{347} & \num{1} & \num{4} \\
    LeetCode      & \num{81} & \num{7} & \num{18} & \num{81} & \num{7} & \num{18}\\
    \cdashlinelr{1-7}
    GeeksForGeeks      & \num{3753} & \num{268} & \num{995} & \num{3753} & \num{268} & \num{995}\\
    Project Euler & \num{110} & \num{11} & \num{41} & \num{110} & \num{11} & \num{41}\\
    \midrule\midrule
    Total     & \num{7009} & \num{443} & \num{1746} & \num{55179} & \num{443} & \num{1746}\\
    \bottomrule
  \end{tabular}
  \end{adjustbox}
\end{table}

\section{The AVATAR-TC Dataset}
\label{sec4:avatartc}

This paper introduces a new dataset \textbf{AVATAR-TC} (built on top of the AVATAR~\cite{ahmad-2021-avatar}) that has pairs of whole-programs in Java and Python (a statically- and dynamically-typed language, with different syntactic styles), each accompanied by human-written test-cases (TCs). Based on what we know, AVATAR-TC is the first such large-scale dataset where code compilability (syntactic correctness) is ensured, and code pairs have undergone thorough testing w.r.t. human-written TCs. 

Note that these TCs are not used in training any of the \cotran variants (which rely on automated unit test generation); in this paper, we use them to evaluate translators with FEqAcc (Section~\ref{subsec4:Metrics}).

\noindent{\bf Data Source.} Similar to AVATAR, we gather a collection of code pairs written in Java and Python by scraping five \textit{competitive coding websites} that host regular contests: Aizu~\cite{aizu}, AtCoder~\cite{atcoder}, Codeforces~\cite{codeforces}, G-CodeJam~\cite{codejam}, LeetCode~\cite{leetcode} and two \textit{coding platforms}: GeeksForGeeks~\cite{geeksforgeeks}, Project Euler~\cite{projecteuler}. These serve as great resources for mining code solutions of a problem statement across multiple languages. Also, a participant's code gets checked on multiple test-cases (TCs) curated by the problem-setter. For AVATAR-TC, we web-crawled these data sources and collected such human-written TCs to complement each problem statement.

\noindent{\bf Data Cleaning.} Several code pairs in AVATAR dataset did not compile and/or pass our collected TCs. Consequently in AVATAR-TC, we preprocessed codes afresh from their sources. Utilizing the \texttt{javalang}~\cite{javalang} and \texttt{tokenize}~\cite{py-tokenize} modules, they were parsed into code-specific tokens. We manually corrected minor faults in code pairs, that did not match the expected output when provided with the TC inputs. Code pairs with major issues were discarded. Our criteria for output matching include case insensitivity, whitespace removal, punctuation disregard (unless significant to the output), and normalization of numeric or floating-point values to a common representation.

\begin{table*}[!t]
\caption{\textbf{Code Translation Results:} Comparison of \cotran against 14 other tools for Java-Python (J2P) and Python-Java (P2J) translation. (In each column, the highest value is marked in \textbf{bold}, second-highest \underline{underlined}.)}
\label{table:results_translation}
\setlength\tabcolsep{4pt}
\centering
\renewcommand{\arraystretch}{1.15}
\begin{adjustbox}{max width=1\textwidth}
\begin{tabular}
{@{}cl|cccccc|cccccc@{}}
\toprule
{\multirow{2}{*}{Method / Tool}}   &   {\multirow{2}{*}{Model}}   &   \multicolumn{6}{|c}{Java $\to$ Python (J2P)}  &   \multicolumn{6}{|c}{Python $\to$ Java (P2J)} \\
\cmidrule(lr){3-8} \cmidrule(lr){9-14}
  &     &   \textit{FEqAcc}   &   \textit{CompAcc}   &   $\text{errPos}_{\nth{1}}$   &   CodeBLEU   &   BLEU   &   \phantom{W}EM\phantom{W}   &   \textit{FEqAcc}   &   \textit{CompAcc}   &   $\text{errPos}_{\nth{1}}$  &   CodeBLEU   &   BLEU   &   \phantom{W}EM\phantom{W} \\
\midrule
{\multirow{3}{*}{Transpilers}} &   java2python~\cite{java2python}   &  3.32  &  41.46  &  28.62  &  20.31  &  17.54  &  0  &   -   &   -   &   -  &   -   &   -   &   - \\
  &   TSS CodeConv~\cite{tangiblesoftsoln}   &  0.46  &  58.30  &  54.26  &  41.87  &  24.44  &  0  &   -   &   -   &   -  &   -   &   -   &   - \\
  &   py2java~\cite{py2java}   &   -   &   -   &   -   &   -   &   -   &   -   &  0  &  0  &  1.61  &  41.56  &  48.59  &  0\\
\noalign{\vskip 0.2ex}\hdashline\noalign{\vskip 0.4ex}
{\multirow{3}{*}{\makecell{Recent \\competing tools\\\textit{(unsupervised trng.)}}}}   &   TransCoder~\cite{roziere2020unsupervised}   &  0.46  &  88.09  &  63.57  &  35.07  &  32.07  &  0  &  0  &  0  &  4.57  &  35.02  &  35.06  &  0\\
  &   TransCoder-DOBF~\cite{lachaux2021dobf}   &  0.46  &  63.00  &  47.10  &  39.98  &  33.84  &  0  &  0  &  0  &  3.11  &  33.33  &  32.72  &  0\\
  &   TransCoder-ST~\cite{roziere2022leveraging}   &  0.46  &  91.58  &  74.68  &  40.04  &  37.30  &  0  &  0  &  0  &  4.67  &  29.88  &  28.15  &  0\\
\noalign{\vskip 0.2ex}\hdashline\noalign{\vskip 0.4ex}
ChatGPT   &   \makecell[l]{GPT-3.5-turbo~\cite{chatgpt}}   &   {\textbf{76.06}}   &   {95.36}   &   {90.88}    &   {52.11}   &   {53.19}   &   {0.29}   &   {21.65}   &   {24.97}   &   {30.86}    &   {54.08}   &   {55.58}   &   {0} \\
\noalign{\vskip 0.2ex}\hdashline\noalign{\vskip 0.4ex}
{\multirow{7}{*}{\makecell{Recent \\competing tools\\(\textit{supervised trng.}\\\textit{on AVATAR-TC)}}}}   &   CodeBERT~\cite{feng2020codebert}   &  12.31  &  84.77  &  79.57  &  46.00  &  48.10  &  0.46  &  0.74  &   \textbf{96.79}   &   \textbf{99.51}   &  26.10  &  19.62  &  0\\
  &   GraphCodeBERT~\cite{guo2020graphcodebert}   &  10.88  &  85.05  &  79.78  &  45.53  &  47.26  &  0.57  &  0.46  &  \underline{89.75}  &  \underline{98.05}  &  23.72  &  16.21  &  0\\
  &   CodeGPT~\cite{lu2021codexglue}   &  24.86  &  78.92  &  89.21  &  38.38  &  38.64  &  1.49  &  13.40  &  45.13  &  94.50  &  40.51  &  37.96  &  0.52\\
  &   CodeGPT-adapted~\cite{lu2021codexglue}   &  24.17  &  76.75  &  89.31  &  36.84  &  37.36  &  1.55  &  20.50  &  52.00  &  97.60  &  41.46  &  38.15  &  1.03\\
  &   PLBART-base~\cite{ahmad2021unified}   &  38.55  &  91.47  &  90.79  &  54.77  &  59.34  &  1.32  &  38.26  &  75.77  &  96.64  &  55.96  &  59.24  &  0.97\\
  &   CodeT5-base~\cite{wang2021codet5}   &  40.95  &  92.84  &  \textbf{93.76}  &  55.34  &   {60.03}   &  2.41  &  33.79  &  68.84  &  98.02  &  57.64  &  60.16  &  0.86\\
  &   PPOCoder~\cite{shojaee2023execution}   &  44.27  &  93.47  &  91.44  &  55.16  &  59.51  &  1.89  &   {37.11}   &   {59.62}   &   {96.77}    &   {55.04}   &   {58.52}   &   {0.52} \\
\cmidrule(lr){1-14}
Our tool   &   \makecell[l]{\textbf{\cotran (baseline)}}   &  44.52  &  96.12  &  92.07  &  55.44  &  58.71  &  2.11  &  40.41  &  73.63  &  92.16  &   \textbf{59.11}   &   {61.12}   &  \textbf{1.66}\\
\noalign{\vskip 0.2ex}\hdashline\noalign{\vskip 0.4ex}
{\multirow{2}{*}{\makecell{Our tool with \\CF only}}}  &   \makecell[l]{\cotran + CF (RL-based fine-tuning)}   &   {47.02}   &   {96.56}   &   {91.58}    &   {56.10}   &   {60.59}   &   \underline{2.23}   &   {42.78}   &   {74.80}   &   {96.91}    &   {58.55}   &   \underline{61.26}   &   \underline{1.60} \\
  &   \makecell[l]{\textbf{\cotran + CF (RL+SFT interleaved)}}   &   {49.83}   &   \underline{96.79}   &   {92.08}    &   {56.07}   &   \underline{60.61}   &   \underline{2.23}   &   \underline{45.93}   &   {75.77}   &   {96.89}    &   {58.28}   &   {61.21}   &   \underline{1.60} \\
\noalign{\vskip 0.2ex}\hdashline\noalign{\vskip 0.4ex}
{\multirow{2}{*}{\makecell{Our tool (b2b)\\ with CF, SF}}}   &   \makecell[l]{\cotran + CF + SF (RL-based fine-tuning)}   &   {50.45}   &   \underline{96.79}   &   {92.15}    &   \underline{56.17}   &   {60.60}   &   \underline{2.23}   &   {43.92}   &   {75.14}   &   {96.93}    &   \underline{58.59}   &   \textbf{61.28}   &   \underline{1.60} \\
  &   \makecell[l]{\textbf{\cotran + CF + SF (RL+SFT interleaved)}}   &   \underline{53.89}   &   {\textbf{97.14}}   &   \underline{92.73}    &   {\textbf{56.24}}   &   {\textbf{60.69}}   &   \textbf{2.29}   &   \textbf{48.68}   &   {76.98}   &   {96.93}    &   {58.38}   &   {61.19}   &   \underline{1.60} \\
\bottomrule
\end{tabular}
\end{adjustbox}
\end{table*}

\noindent{\bf Statistics of AVATAR-TC dataset.} The train/validation/test partitioning of problem statements is kept the same as AVATAR, except removal of some pairs during data cleaning. This resulted in \num{57368} Java-Python pairs at a train\,:\,validation\,:\,test ratio of $76:5:19$, across \num{9198} problems. For the train split, at most 25 pairs correspond to one problem, while for validation and test, there is a unique one-one pair-problem mapping. To ensure \textit{out-of-distribution testing}, no problem overlaps across splits. Refer to Table~\ref{table:avatarModeifiedStats} for AVATAR-TC statistics.

\section{Experiments}
\label{sec4:experiments}

\subsection{Experimental Setup and Competing Tools}
\label{subsec4:implementation}
\noindent \textbf{\cotran (baseline)} refers to an LLM fine-tuned without feedback loops (refer Figure~\ref{fig:CoTranBaseline}). We use the pre-trained CodeT5-base~\cite{wang2021codet5} architecture from Huggingface~\cite{huggngface} and fine-tune it with CE loss, employing the proposed keyword-based tokenizer (kw-Tok). The maximum length for the source and target sequences is set at $512$. For additional design choices adopted during implementing \cotran, please refer to Appendix A.2. The baseline \cotran and its variants are compared against (See Table~\ref{table:results_translation}): (a) three \textit{human-written transpilers}, (b) three SoTA \textit{LLM-based unsupervised translation tools} (trained on function pairs from ${\sim}2.5$M open-sourced repositories of the GitHub dataset from Google BigQuery Public Datasets), (c) \textit{ChatGPT}~\cite{chatgpt} and, (d) seven \textit{LLM-based supervised translation tools}. All the tools above are compared on the same set of \num{1746} whole-programs from AVATAR-TC (Test). Additionally, each of the supervised tools and \cotran variants are fine-tuned on AVATAR-TC (Train). For ChatGPT, we use OpenAI API to access the \texttt{gpt-3.5-turbo-0301} model. This version of ChatGPT has a knowledge cutoff of March 1, 2023, which predates the public release of AVATAR-TC on GitHub. 
This minimizes the risk of AVATAR-TC (Test) pairs being included in ChatGPT’s training data, ensuring a fair evaluation. Following a standardized protocol~\cite{yan2023codetransocean} for ChatGPT, we use "\texttt{Translate [$S$] to [$T$]:[$\mathbf{s}_{code}$]\textbackslash n Do not return anything other than the translated code.}" as the prompt. Temperature and top\_p are set at $0$ (this ensures reproducibility and does not notably alter the translation performance). 

\subsection{Evaluation Metrics}
\label{subsec4:Metrics}

We evaluate our method using \textit{greedy decoding} that considers only the top translation with the highest log probability. The different metrics for evaluating code translation quality are:

\noindent{\bf BLEU, CodeBLEU score.} BLEU~\cite{papineni2002bleu} computes the `closeness' with the reference translation through n-gram overlaps. CodeBLEU~\cite{ren2020codebleu} additionally checks weighted n-gram match, syntactic AST match (SM) and semantic data-flow match (DM). Both range in $[0,100]$.

\noindent{\bf Exact String Match (EM).} EM is the percentage of generated codes that exactly match the reference translation. It can be low even if the generated codes are compilable and functionally equivalent.


\noindent{\bf Compilation Accuracy (\textit{CompAcc}) and Functional-Equivalence Accuracy (\textit{FEqAcc}).} \textit{CompAcc} is defined as the percentage of generated translations that compile correctly. \textit{FEqAcc} is the percentage of generated translations that are IO equivalent to the source-language code w.r.t. a set of human-written test-cases. 

Additionally, we propose three new quality measures, namely:

\noindent{\bf Average First Error Position ($\mathbf{errPos_{\nth{1}}}$).} $\text{errPos}_{\nth{1}}$ is a fine-grained version of \textit{CompAcc}, relating to the closeness of the translations from a perfect compilation. Averaged over all translations on test set, $\text{errPos}_{\nth{1}}(\,\widehat{\mathbf{t}}\,)$ calculates the position of the first token responsible for a syntactic error in $\widehat{\mathbf{t}}$, normalized by $|\,\widehat{\mathbf{t}}\,|$. It is computed by \texttt{pylint} and \texttt{javac} for Python and Java respectively. Let's consider that a translator achieves $\text{errPos}_{\nth{1}}=e\%$. This implies that, on average, the first compilation error (if any) is located within the last $(100-e)\%$ portion of the generated translations. As $e$ approaches 100, the human developer only needs to inspect a small section of each translated code to rectify the error, thereby facilitating ease of manual debugging. 

\noindent{\bf Average \#Errors per Code (EpC).} EpC is the average count of compilation (syntactic) errors per translated code. Zero errors indicate full syntactic correctness. A higher count implies that substantial effort from a human end-user is required for rectification.

\noindent{\bf Ratio of FEqAcc and CompAcc ($\frac{\text{f}}{\text{c}}$rate).} $\frac{\text{f}}{\text{c}}$rate is the percentage rate at which a translator generates functionally-equivalent codes compared to the compilable ones. In a real-world deployment, checking equivalence of translator-generated codes is infeasible. Thus, it is beneficial when a syntactically correct code produced by a translator implies functional equivalence i.e., ideally $\frac{\text{f}}{\text{c}}$rate tends to $100\%$. $\frac{\text{f}}{\text{c}}$rate is computed as the percentage ratio of FEqAcc and CompAcc. 




\subsection{Analysis of Empirical Results and Ablation Study}
\label{subsec4:Results}

In Table~\ref{table:results_translation}, we compare \cotran against 14 competing tools for J2P and P2J translation. PLBART-base, CodeT5-base, and PPOCoder perform best among the competitors, while transpilers underperform (refer Appendix A.4). ChatGPT exhibits weak performance in P2J with a significant shortfall in FEqAcc ($-27.03\%$), CompAcc ($-52.01\%$), and $\text{errPos}_{\nth{1}}$ ($-66.07\%$) compared to the best CoTran method. A low $\text{errPos}_{\nth{1}}$ indicates that ChatGPT-generated translations are hard to debug. But, in J2P, ChatGPT leads with a FEqAcc of $76.06\%$. Given that \cotran is built on CodeT5-base with ${\sim}220$M parameters, we believe that fine-tuning a model as large as ChatGPT (rumored to have 100B+ parameters) using our method would outperform ChatGPT in J2P across all metrics. However, while ChatGPT supports fine-tuning, it is closed-source and its interface is not yet tailored to receive symbolic feedback from tools such as compilers, testers, and solvers. Hereafter, \textbf{we report \cotran improvements w.r.t. CodeT5-base}.



\vspace{2mm}
\noindent \textbf{Hypothesis:} \textit{LLMs are good at code translation. Having said that, incorporating compiler and symexec feedback (CF, SF) during fine-tuning significantly improves its capability of producing compilable and functionally equivalent translations.}

\noindent Our results validate this hypothesis when evaluated across the diverse AVATAR-TC dataset. In the process, we have made several findings:

\vspace{2mm}
\noindent \textbf{Finding 1:} \textit{RL turned out to be much more effective than the SFT schemes we tried, for incorporating feedback during fine-tuning.}

\noindent We considered two \textit{non-RL schemes} of SFT-based LLM training, combining CE loss and CF, SF. However, they underperform compared to the RL-based methods. Please refer to Appendix A.3 for more details.

\begin{figure}[!t]
    \centering
    \pgfplotsset{width=\columnwidth,height=0.25\textwidth,compat=1.9}\begin{tikzpicture}
\footnotesize
\begin{axis}[
xlabel={$\left|\,\widehat{\mathbf{t}}\,\right|/\left|\mathbf{t}\right|$},
ylabel={Reward Value},
xmin=-0.02, xmax=1.02,
ymin=-1.1, ymax=2.1,
xtick={0,0.1,0.2,0.3,0.4,0.5,0.6,0.7,0.8,0.9,1.0},
ytick={-1.0,-0.5,0,0.5,1,1.5,2},
xticklabel style={rotate=60},
legend columns=2,
legend style={at={(0,1)},anchor=north west,font=\mysizeSub,row sep=0.2pt},
ymajorgrids=true,
grid style=dashed
]

\addlegendentry{$\omega_{CF}(\mathbf{t}, \widehat{\mathbf{t}}\,)$}
\addplot[
color=red,
mark=diamond,
mark size=1.5pt,
line width=0.7pt
]
coordinates {
(0.0123456790123457,0.0063)
(0.0246913580246914,0.00951)
(0.037037037037037,0.01209)
(0.0493827160493827,0.01455)
(0.0617283950617284,0.01706)
(0.0740740740740741,0.01973)
(0.0864197530864197,0.0226)
(0.0987654320987654,0.02571)
(0.111111111111111,0.02911)
(0.123456790123457,0.03281)
(0.135802469135802,0.03686)
(0.148148148148148,0.04128)
(0.160493827160494,0.04609)
(0.172839506172839,0.05134)
(0.185185185185185,0.05704)
(0.197530864197531,0.06323)
(0.209876543209877,0.06993)
(0.222222222222222,0.07718)
(0.234567901234568,0.085)
(0.246913580246914,0.09342)
(0.259259259259259,0.10247)
(0.271604938271605,0.11217)
(0.283950617283951,0.12254)
(0.296296296296296,0.13362)
(0.308641975308642,0.14542)
(0.320987654320988,0.15796)
(0.333333333333333,0.17126)
(0.345679012345679,0.18532)
(0.358024691358025,0.20017)
(0.37037037037037,0.2158)
(0.382716049382716,0.23223)
(0.395061728395062,0.24944)
(0.407407407407407,0.26743)
(0.419753086419753,0.2862)
(0.432098765432099,0.30572)
(0.444444444444444,0.32598)
(0.45679012345679,0.34694)
(0.469135802469136,0.36858)
(0.481481481481481,0.39085)
(0.493827160493827,0.41372)
(0.506172839506173,0.43714)
(0.518518518518518,0.46104)
(0.530864197530864,0.48537)
(0.54320987654321,0.51006)
(0.555555555555556,0.53503)
(0.567901234567901,0.56022)
(0.580246913580247,0.58553)
(0.592592592592593,0.61088)
(0.604938271604938,0.63619)
(0.617283950617284,0.66135)
(0.62962962962963,0.68627)
(0.641975308641975,0.71084)
(0.654320987654321,0.73497)
(0.666666666666667,0.74451)
(0.679012345679012,0.7815)
(0.691358024691358,0.78933)
(0.703703703703704,0.82501)
(0.716049382716049,0.82501)
(0.728395061728395,0.82501)
(0.740740740740741,0.82501)
(0.753086419753086,0.82501)
(0.765432098765432,0.82501)
(0.777777777777778,0.82501)
(0.790123456790123,0.82501)
(0.802469135802469,0.82501)
(0.814814814814815,0.82501)
(0.827160493827161,0.82501)
(0.839506172839506,0.84539)
(0.851851851851852,0.86472)
(0.864197530864197,0.8829)
(0.876543209876543,0.89985)
(0.888888888888889,0.91548)
(0.901234567901235,0.91496)
(0.91358024691358,0.94248)
(0.925925925925926,0.93903)
(0.938271604938272,0.96333)
(0.950617283950617,0.97132)
(0.962962962962963,0.97762)
(0.975308641975309,0.9822)
(0.987654320987654,0.98504)
(1,2)
};

\addlegendentry{SM$(\mathbf{t}, \widehat{\mathbf{t}}\,)$}
\addplot[
color=olive,
mark=o,
mark size=1.5pt,
line width=0.7pt
]
coordinates {
(0.0123456790123457,0)
(0.0246913580246914,0)
(0.037037037037037,0)
(0.0493827160493827,0)
(0.0617283950617284,0.0294)
(0.0740740740740741,0.0294)
(0.0864197530864197,0.0294)
(0.0987654320987654,0.0294)
(0.111111111111111,0.0294)
(0.123456790123457,0.0294)
(0.135802469135802,0.0882)
(0.148148148148148,0.1176)
(0.160493827160494,0.1471)
(0.172839506172839,0.1471)
(0.185185185185185,0.2059)
(0.197530864197531,0.2059)
(0.209876543209877,0.2059)
(0.222222222222222,0.2647)
(0.234567901234568,0.2647)
(0.246913580246914,0.2647)
(0.259259259259259,0.2647)
(0.271604938271605,0.2647)
(0.283950617283951,0.2941)
(0.296296296296296,0.2941)
(0.308641975308642,0.2941)
(0.320987654320988,0.2941)
(0.333333333333333,0.2941)
(0.345679012345679,0.2941)
(0.358024691358025,0.3824)
(0.37037037037037,0.3824)
(0.382716049382716,0.3824)
(0.395061728395062,0.3824)
(0.407407407407407,0.4118)
(0.419753086419753,0.4118)
(0.432098765432099,0.4118)
(0.444444444444444,0.4118)
(0.45679012345679,0.4118)
(0.469135802469136,0.4118)
(0.481481481481481,0.4118)
(0.493827160493827,0.4118)
(0.506172839506173,0.4412)
(0.518518518518518,0.4706)
(0.530864197530864,0.4706)
(0.54320987654321,0.4706)
(0.555555555555556,0.4706)
(0.567901234567901,0.4706)
(0.580246913580247,0.4706)
(0.592592592592593,0.4706)
(0.604938271604938,0.5294)
(0.617283950617284,0.5588)
(0.62962962962963,0.6176)
(0.641975308641975,0.6176)
(0.654320987654321,0.6176)
(0.666666666666667,0.6176)
(0.679012345679012,0.6176)
(0.691358024691358,0.6176)
(0.703703703703704,0.6765)
(0.716049382716049,0.6765)
(0.728395061728395,0.6765)
(0.740740740740741,0.6765)
(0.753086419753086,0.6765)
(0.765432098765432,0.6765)
(0.777777777777778,0.6765)
(0.790123456790123,0.6765)
(0.802469135802469,0.6765)
(0.814814814814815,0.6765)
(0.827160493827161,0.6765)
(0.839506172839506,0.6765)
(0.851851851851852,0.6765)
(0.864197530864197,0.6765)
(0.876543209876543,0.6765)
(0.888888888888889,0.6765)
(0.901234567901235,0.6765)
(0.91358024691358,0.6765)
(0.925925925925926,0.6765)
(0.938271604938272,0.6765)
(0.950617283950617,0.6765)
(0.962962962962963,0.8235)
(0.975308641975309,0.8529)
(0.987654320987654,0.9118)
(1,1)
};

\addlegendentry{DM$(\mathbf{t}, \widehat{\mathbf{t}}\,)$}
\addplot[
color=blue,
mark=o,
mark size=1.5pt,
line width=0.7pt
]
coordinates {
(0.0123456790123457,0)
(0.0246913580246914,0)
(0.037037037037037,0)
(0.0493827160493827,0)
(0.0617283950617284,0)
(0.0740740740740741,0)
(0.0864197530864197,0)
(0.0987654320987654,0)
(0.111111111111111,0)
(0.123456790123457,0)
(0.135802469135802,0)
(0.148148148148148,0)
(0.160493827160494,0)
(0.172839506172839,0)
(0.185185185185185,0)
(0.197530864197531,0)
(0.209876543209877,0)
(0.222222222222222,0.1429)
(0.234567901234568,0.1429)
(0.246913580246914,0.1429)
(0.259259259259259,0.1429)
(0.271604938271605,0.2857)
(0.283950617283951,0.2857)
(0.296296296296296,0.2857)
(0.308641975308642,0.2857)
(0.320987654320988,0.2857)
(0.333333333333333,0.3571)
(0.345679012345679,0.3571)
(0.358024691358025,0.3571)
(0.37037037037037,0.3571)
(0.382716049382716,0.4286)
(0.395061728395062,0.4286)
(0.407407407407407,0.5)
(0.419753086419753,0.5)
(0.432098765432099,0.5)
(0.444444444444444,0.5)
(0.45679012345679,0.5)
(0.469135802469136,0.5)
(0.481481481481481,0.5714)
(0.493827160493827,0.5714)
(0.506172839506173,0.7143)
(0.518518518518518,0.7143)
(0.530864197530864,0.7857)
(0.54320987654321,0.7857)
(0.555555555555556,0.7857)
(0.567901234567901,0.7857)
(0.580246913580247,0.8571)
(0.592592592592593,0.7857)
(0.604938271604938,1)
(0.617283950617284,1)
(0.62962962962963,1)
(0.641975308641975,1)
(0.654320987654321,1)
(0.666666666666667,1)
(0.679012345679012,1)
(0.691358024691358,1)
(0.703703703703704,1)
(0.716049382716049,1)
(0.728395061728395,1)
(0.740740740740741,1)
(0.753086419753086,1)
(0.765432098765432,1)
(0.777777777777778,1)
(0.790123456790123,1)
(0.802469135802469,1)
(0.814814814814815,1)
(0.827160493827161,1)
(0.839506172839506,1)
(0.851851851851852,1)
(0.864197530864197,1)
(0.876543209876543,1)
(0.888888888888889,1)
(0.901234567901235,1)
(0.91358024691358,1)
(0.925925925925926,1)
(0.938271604938272,1)
(0.950617283950617,1)
(0.962962962962963,1)
(0.975308641975309,1)
(0.987654320987654,1)
(1,1)
};

\addlegendentry{compiler${}_{\text{Bool}}(\,\widehat{\mathbf{t}}\,)$}
\addplot[
color=OliveGreen,
mark=diamond,
mark size=1.5pt,
line width=0.7pt
]
coordinates {
(0.0123456790123457,-1)
(0.0246913580246914,-1)
(0.037037037037037,-1)
(0.0493827160493827,-1)
(0.0617283950617284,-1)
(0.0740740740740741,-1)
(0.0864197530864197,-1)
(0.0987654320987654,-1)
(0.111111111111111,-1)
(0.123456790123457,-1)
(0.135802469135802,-1)
(0.148148148148148,-1)
(0.160493827160494,-1)
(0.172839506172839,-1)
(0.185185185185185,-1)
(0.197530864197531,-1)
(0.209876543209877,-1)
(0.222222222222222,-1)
(0.234567901234568,-1)
(0.246913580246914,-1)
(0.259259259259259,-1)
(0.271604938271605,-1)
(0.283950617283951,-1)
(0.296296296296296,-1)
(0.308641975308642,-1)
(0.320987654320988,-1)
(0.333333333333333,-1)
(0.345679012345679,-1)
(0.358024691358025,-1)
(0.37037037037037,-1)
(0.382716049382716,-1)
(0.395061728395062,-1)
(0.407407407407407,-1)
(0.419753086419753,-1)
(0.432098765432099,-1)
(0.444444444444444,-1)
(0.45679012345679,-1)
(0.469135802469136,-1)
(0.481481481481481,-1)
(0.493827160493827,-1)
(0.506172839506173,-1)
(0.518518518518518,-1)
(0.530864197530864,-1)
(0.54320987654321,-1)
(0.555555555555556,-1)
(0.567901234567901,-1)
(0.580246913580247,-1)
(0.592592592592593,-1)
(0.604938271604938,-1)
(0.617283950617284,-1)
(0.62962962962963,-1)
(0.641975308641975,-1)
(0.654320987654321,-1)
(0.666666666666667,-1)
(0.679012345679012,-1)
(0.691358024691358,-1)
(0.703703703703704,-1)
(0.716049382716049,-1)
(0.728395061728395,-1)
(0.740740740740741,-1)
(0.753086419753086,-1)
(0.765432098765432,-1)
(0.777777777777778,-1)
(0.790123456790123,-1)
(0.802469135802469,-1)
(0.814814814814815,-1)
(0.827160493827161,-1)
(0.839506172839506,-1)
(0.851851851851852,-1)
(0.864197530864197,-1)
(0.876543209876543,-1)
(0.888888888888889,-1)
(0.901234567901235,-1)
(0.91358024691358,-1)
(0.925925925925926,-1)
(0.938271604938272,-1)
(0.950617283950617,-1)
(0.962962962962963,-1)
(0.975308641975309,-1)
(0.987654320987654,-1)
(1,1)
};

\end{axis}
\end{tikzpicture}
    \caption{\textbf{Reward Analysis:} Plot of different rewards for RL}
    \vspace{5mm}
    \label{fig:rewardsPlot}
\end{figure}

\noindent \textbf{Finding 2:} \textit{For RL reward, a Boolean feedback from compiler and other existing feedback are not as effective as ours.}

\noindent For RL-based optimization of LLMs, it is essential to fabricate a good reward function. CompCoder~\cite{wang2022compilable} attempts RL-based code generation using a Boolean feedback compiler${}_{\text{Bool}}(\,\widehat{\mathbf{t}}\,)$, that returns $-1$ or $+1$. PPOCoder~\cite{shojaee2023execution} uses the sum of compiler${}_{\text{Bool}}(\,\widehat{\mathbf{t}}\,)$, syntactic match score SM$(\mathbf{t}, \widehat{\mathbf{t}}\,)$ and dataflow match score DM$(\mathbf{t}, \widehat{\mathbf{t}}\,)$ as the RL reward. We hypothesize that even though these functions are good tools to compare $\widehat{\mathbf{t}}$ with $\mathbf{t}$, they are not the best when it comes to an RL reward.

In Figure~\ref{fig:rewardsPlot}, $\mathbf{t}$ is a tokenized Java code for reversing an integer using for-loop, while $\widehat{\mathbf{t}}$ is its truncated version. We vary $|\,\widehat{\mathbf{t}}\,|/\left|\mathbf{t}\right|$ from 0 to 1 and plot the respective reward. For a significant portion of the $x$-axis, especially when $\widehat{\mathbf{t}}$ is near empty or is almost same as $\mathbf{t}$, values of compiler${}_{\text{Bool}}(\,\widehat{\mathbf{t}}\,)$, SM$(\mathbf{t}, \widehat{\mathbf{t}}\,)$ and DM$(\mathbf{t}, \widehat{\mathbf{t}}\,)$ remain constant. So, CompCoder and PPOCoder offer the same RL reward for several closely-related translations. This limits the RL agent's ability to gauge improvement. In contrast, our proposed reward $\omega_{CF}(\mathbf{t}, \widehat{\mathbf{t}}\,)$ better detects small changes, guiding the RL agent towards smaller goals and improving the overall compilation accuracy. Thus, \cotran + CF (RL only) outperforms PPOCoder, despite both using CodeT5-base.

\noindent \textbf{Finding 3:} \textit{Interleaving RL and SFT improves the LLM's performance, compared to an RL-only fine-tuning approach.}

\noindent To prevent RL-based fine-tuning from deviating the back-to-back LLMs from their CE loss objective, we occasionally interleave it with supervised fine-tuning (SFT). This improves the overall translation performance. In order to incorporate CF (\textbf{\cotran + CF}), we fine-tune the baseline model in an RL setting to maximize CF as a reward (refer Figure~\ref{fig:CoTranWithCF}). This approach improves J2P and P2J translation by $+{6.07}\%$, $+{8.99}\%$ in FEqAcc, and $+{3.72}\%$, $+{5.96}\%$ in CompAcc, respectively. The RL+SFT interleaved training boosts these gains to $+8.88\%$, $+12.14\%$ in FEqAcc and $+3.95\%$, $+6.93\%$ in CompAcc. Note that for \cotran + CF, the interleaved training is similar to Algorithm~\ref{algo:b2bTrain} with the difference that $\omega_{\text{SF}}$ is not considered (there are no b2b LLMs, only the forward LLM). When both CF and SF are incorporated (\textbf{\cotran + CF + SF}) using RL with b2b LLMs (Figure~\ref{fig:CoTranWithCFSF}), the improvements are $+{9.50}\%$, $+{10.13}\%$ in FEqAcc and $+{3.95}\%$, $+{6.30}\%$ in CompAcc. Interleaving RL+SFT (Figure~\ref{fig:CoTranInterleavedLoop} and Algorithm~\ref{algo:b2bTrain}) further increases these to $+12.94\%$, $+14.89\%$ in FEqAcc and $+4.30\%$, $+8.14\%$ in CompAcc. Additionally, \cotran + CF + SF achieves $92.73\%$ and $96.93\%$ on \textit{errPos}${}_{\nth{1}}$ for J2P and P2J, thus signifying easy-to-debug translations.

\noindent \textbf{Finding 4:} \textit{Adding keywords to tokenizer vocabulary by kw-Tok improves the code translation performance.}

\noindent For J2P and P2J translation respectively, \textit{kw-Tok} itself accounts for $+3.57\%$, $+6.62\%$ increase on FEqAcc and, $+3.28\%$, $+4.79\%$  increase on CompAcc (\cotran baseline v/s CodeT5-base).

\noindent \textbf{Finding 5:} \textit{Incorporating compiler and symexec feedback during fine-tuning reduces the count of syntactic errors per translated code.}

\noindent In Figure~\ref{fig:numErrorsPlot}, we compare the number of compilation (syntactic) errors per code (EpC) produced in Python-to-Java (P2J) translation on AVATAR-TC (Test). ChatGPT exhibits significantly poorer performance with an EpC of $10.98\pm 9.67$ (i.e., mean $\pm$ stdev). CodeT5-base reduces this to $0.95\pm 2.58$. Our \cotran + CF + SF (RL+SFT) further lowers the EpC value to $0.71\pm 2.15$. This indicates that the proposed feedback results in producing more easy-to-debug translations.

\begin{figure}[t]
    \centering
    \pgfplotsset{width=\columnwidth,height=0.2\textwidth,compat=1.9}\input{Figures/numErrorsPlot.tex}
    \caption{\textbf{Count of compilation errors per code (EpC)}: \#Errors per translated code in P2J translation, sorted in the ascending order}
    \vspace{5mm}
    \label{fig:numErrorsPlot}
\end{figure}

\noindent \textbf{Finding 6:} \textit{
Integrating compiler and symexec feedback during fine-tuning markedly improves $\frac{\text{f}}{\text{c}}$rate metric across diverse codebases.}

\noindent In Figure~\ref{fig:performanceByDataset}, we compare the P2J performance of four SoTA tools and three \cotran-based methods over the different sub-datasets of AVATAR-TC (Test). CodeBERT demonstrates the highest rate of compilable translations (with over 90\% CompAcc across most sub-datasets). However, the resultant translations lack meaningfulness, with only around 1\% of them being functionally equivalent. In contrast, CodeGPT, PLBART-base, and CodeT5-base yield functionally equivalent codes at rates of 29.69\%, 50.49\%, and 49.08\% respectively, relative to the number of compilable translations they generate across all sub-datasets. This $\frac{\text{f}}{\text{c}}$rate increases to 63.24\% for  \cotran + CF + SF (RL+SFT). Our method consistently \textit{ranks among the top two} in $\frac{\text{f}}{\text{c}}$rate for each sub-dataset. Note that G-CodeJam is a small sub-dataset, and all tools exhibit similarly poor performance when translating P2J.

\noindent \textbf{Finding 7:} \textit{It is easier for a code translator to achieve good performance if the source language is statically-typed.}

\noindent Java is a statically-typed language, while Python is dynamically-typed. When translating J2P, Java's explicit variable-type declarations are ignored as they are redundant in Python.
 Conversely, in Python, variable-types are not explicitly declared, requiring the translator to infer them during P2J translation. Even for humans experienced in both languages, deducing types for variables in Python can be challenging and demand repeated engrossed mental evaluations. Consequently, learning P2J is much more challenging than J2P. Thus, the source language being statically-typed makes learning code translation easier.

\noindent \textbf{Finding 8:} \textit{Function-to-function translation is not sufficient for whole-program translation tasks.}

\noindent The TransCoder-based unsupervised tools are trained on function-level translations. As per Roziere et al.~\cite{roziere2020unsupervised}, this keeps training batches shorter and unit test-based model evaluation simpler. But consequently, these tools cannot efficiently generate compilable, functionally-equivalent translations for whole-programs. For instance, in J2P, all three TransCoder-based tools fail drastically in FEqAcc (0.46\%). Similarly, for P2J, no (0\%) translated code is compilable.

\begin{figure}[t]
    \centering
    \pgfplotsset{width=1\columnwidth,height=0.24\textwidth,compat=1.9}\input{Figures/performanceByDataset.tex}
    \caption{\textbf{Sub-dataset wise performance w.r.t. $\frac{\text{f}}{\text{c}}$rate:} Comparison of P2J translation, across the 7 sub-datasets of AVATAR-TC (Test) 
    }
    \vspace{5mm}
    \label{fig:performanceByDataset}
\end{figure}

\vspace{1.2mm}
\noindent \textbf{Benefits of automatic test-case generation (TCgen) during training.} During fine-tuning $S \to T \to S$ back-to-back LLMs (Section~\ref{subsec3:training}), we automatically generate unit tests on $\mathbf{s}$ and test them on $\widehat{\mathbf{s}}$. Note that equivalence checking of programs is an undecidable problem. Our idea is to assess the \textit{inequivalence} of $\mathbf{s}$ and $\widehat{\mathbf{s}}$ (detectable via sufficient testing) w.r.t. a test suite produced by TCgen tool and thereby, to calculate symexec feedback (SF) for the LLM during fine-tuning. To our knowledge, no existing LLM-based code translation tool employs \textit{automatic symexec-based TCgen} for functional equivalence checking and feedback. All existing tools like PPOCoder~\cite{shojaee2023execution}, TransMap~\cite{wang2023transmap}, RLTF~\cite{liu2023rltf}, CodeRL~\cite{le2022coderl} use \textit{human-written} test-cases for fixing LLM-generated codes. In contrast, \cotran does not require anything extra than what is required for a CE loss-based supervised fine-tuning of LLM i.e., a dataset of $S, T$ code pairs. Also, compared to human-written test-cases, symexec-based testing has a higher probability of covering all linearly independent control-flow paths. 

\noindent \textbf{Generalizability regarding choice of TCgen tool.} We want to emphasize that our approach is entirely agnostic to any specific TCgen tool, including Symflower. The selection of Symflower is driven by the fact that we tested \cotran with Java as $S$ in the $S \to T \to S$ back-to-back training loop, and Symflower stands out as a commercial industrial-strength symexec engine for Java. Its efficiency and reliability make it a preferred choice. MLB, JBMC, and GDart (the top 3 verification tools for Java in SV-COMP 2024~\cite{beyer2024state}) can also be considered as potential alternatives to Symflower. In fact, similar efficient TCgen tools (e.g., EvoSuite~\cite{fraser2011evosuite} for Java, KLOVER~\cite{li2011klover} for C++, Pynguin~\cite{lukasczyk2022pynguin} for Python) are available for most popular languages. If not, LLVM-based TCgen tools (e.g., KLEE) offer an alternative, as many languages can be readily translated into LLVM IR.

\vspace{1.2mm}
\noindent \textbf{Reproducibility and Supplementary.} The AVATAR-TC dataset and all our code can be accessed at  \url{https://github.com/PrithwishJana/CoTran}. The supplementary material (Appendix) is provided at the end of this paper.

\vspace{-2mm}
\section{Conclusion and Future Work}
\label{sec5:conclusion}

In this paper, we present an LLM-based code translation method (\cotran) that incorporates feedback from compiler and symexec-based solver during fine-tuning. The paper showcases the power of symbolic feedback in making code translation LLMs more accurate. Another key insight is the efficacy of fine-tuning LLMs with fine-grained feedback, pinpointing proximity to an ideal solution, rather than simple Boolean directives like yes/no. Our results show that CF and SF, especially with RL+SFT interleaved training, significantly improve code translation quality. \cotran outperforms state-of-the-art tools in producing compilable and functionally equivalent codes. We plan to extend \cotran to translate legacy code to modern languages.


\bibliography{mybibfile}

\clearpage 
\appendix
\section{Appendix}

\subsection{Reproducibility}

Our code and the AVATAR-TC dataset are available at  \url{https://github.com/PrithwishJana/CoTran}. The repository includes Python-to-Java (P2J) and Java-to-Python (J2P) translations generated by the state-of-the-art translators and \cotran variants (tabulated in Table~\ref{table:results_translation_fullAppendix}). We have also made it user-friendly with a README file in the root folder that outlines the folder structure, library dependencies, and instructions for running the code.

\subsection{\cotran Implementation and Design Choices}
\label{appndixSec:implementation}

Here, we outline some of the additional design choices we adopted for implementing \cotran. For symbolic execution feedback (SF), Symflower generates JUnit tests for each method (say, named $p$) in the input Java code $\mathbf{s}$. These JUnit tests for $p$ are checked on the corresponding method named $p^*$ of $\widehat{\mathbf{s}}$, such that $p^*=\text{argmax}_{p'\in \;\widehat{\mathbf{s}}}\left(\text{JaccardSimilarity}(p,p')\right)$. 
This accounts for the possibility that methods in $\mathbf{s}$ and $\widehat{\mathbf{s}}$ might have slightly different names but exhibit similar input-output (IO) behavior, making them effectively equivalent. Symflower successfully generated JUnit tests for $\num{5738}$ Java codes in AVATAR-TC (Train) with an average of $6.34$ tests per code. We implemented the training pipeline using Pytorch on a compute node with four NVIDIA V100 GPUs (32GB memory) and six CPU cores per GPU. Low-rank adaptation (LoRA) matrices of the query/value layers use a rank $r$ of $16$ and a scaling factor $\alpha$ of $32$. For optimization of LLM by CE loss, we use the Adam optimizer with a learning rate (lr) of $10^{-4}$. For reinforcement learning (RL)-based optimization of LLMs by PPO algorithm, the output sequence is generated by pure sampling from the LLM distribution, with lr of $1.41\times 10^{-5}$. Further, the parameter-efficient approach of fine-tuning in Algorithm 1 reduces total training time by ${\sim}75\%$ compared to jointly training two LLMs without low-rank optimization.

Regarding the need for a new benchmark suite, most code translation datasets focus on pairs of equivalent snippets or functions in two different languages, rather than whole-programs. Further, they often involve similar languages like Java and C\#, both of which are statically-typed languages with similar syntax. In contrast, the AVATAR-TC dataset includes pairs of whole-programs in Java and Python -- a statistically-typed and a dynamically-typed language with different syntactic styles. Further, to the best of our knowledge, AVATAR-TC is the first large-scale dataset ensuring code compilability (syntactical correctness), and where code pairs have undergone thorough testing with human-written test-cases (TCs).

\subsection{Non-RL methods for integrating CF, SF in training}
\label{appndixSec:nonRLSchemes}

In the main paper, we propose using an RL-based method to integrate compiler feedback (CF) and symbolic execution feedback (SF) into the LLM training process. However, we also explored non-RL methods for incorporating this feedback. Instead of using CF and SF as rewards in the RL framework (Algorithm 1), another approach is to combine CF and SF with the cross-entropy (CE) loss and train the LLM using Supervised Fine-Tuning (SFT). The two alternatives are:

\begin{enumerate}[label=\alph*.]
    \item $\mathbf{\cotran^{+}}$ \textbf{(additive approach):} Minimize a linearly-weighted combination of CE loss and both the feedback.
    
    \item $\mathbf{\cotran^{\times}}$ \textbf{(multiplicative approach):} Minimize a weighted CE loss, where weights for samples in a mini-batch are a combination of both the feedback.
\end{enumerate}

\noindent For $\text{\cotran}^{+}$, with $\alpha_c,\alpha_s\in[0,1]$ as the hyperparameters, we define the combined loss terms for the forward model ($\text{LLM}_{f}$) and the backward model ($\text{LLM}_{b}$) in a back-to-back (b2b) translation pipeline as follows:

\begin{equation}
\resizebox{1.04\columnwidth}{!}{$%
\begin{aligned}
\mathcal{L}_f^{\theta_f}\left(\mathbf{s}, \widehat{\mathbf{s}}, \mathbf{t}, \widehat{\mathbf{t}}\right) &= \begin{bmatrix} 
\mathcal{L}^{\theta_f}_{CE}\left(\mathbf{t}, \widehat{\mathbf{t}}\right) & 
    \begin{bmatrix} 
        \omega_{CF}(\mathbf{t}, \widehat{\mathbf{t}}\,) & \omega_{SF}(\mathbf{s}, \widehat{\mathbf{s}})
    \end{bmatrix}
    \cdot 
    \begin{bmatrix} 
        1-\alpha_s \\ \alpha_s
    \end{bmatrix}
\end{bmatrix}\cdot 
\begin{bmatrix} 
    1-\alpha_c \\ \alpha_c
\end{bmatrix}\\
\mathcal{L}_b^{ \theta_b}\left(\mathbf{s}, \widehat{\mathbf{s}}\right) &= \begin{bmatrix} 
\mathcal{L}^{\theta_b}_{CE}\left(\mathbf{s}, \widehat{\mathbf{s}}\right) & 
    \begin{bmatrix} 
        \omega_{CF}(\mathbf{s}, \widehat{\mathbf{s}}) & \omega_{SF}(\mathbf{s}, \widehat{\mathbf{s}})
    \end{bmatrix}
    \cdot 
    \begin{bmatrix} 
        1-\alpha_s \\ \alpha_s
    \end{bmatrix}
\end{bmatrix}\cdot 
\begin{bmatrix} 
    1-\alpha_c \\ \alpha_c
\end{bmatrix}
\end{aligned}
\label{eq:fwdANDbkwdModelLoss_cotranPlus}
$}
\end{equation}

\noindent For $\text{\cotran}^{\times}$, we train by weighing the CE loss with the reciprocal product of CF and SF. The combined losses for the forward and backward models are defined as follows:

\begin{equation}
\begin{split}
\mathcal{L}_f^{\theta_f}\left(\mathbf{s}, \widehat{\mathbf{s}}, \mathbf{t}, \widehat{\mathbf{t}}\right) &= \mathcal{L}^{\theta_f}_{CE}\left(\mathbf{t}, \widehat{\mathbf{t}}\right)/ \left(\omega_{CF}(\mathbf{t}, \widehat{\mathbf{t}}\,)\times \omega_{SF}(\mathbf{s}, \widehat{\mathbf{s}})\right)\\
\mathcal{L}_b^{\theta_b}\left(\mathbf{s}, \widehat{\mathbf{s}}\right) &= \mathcal{L}^{\theta_b}_{CE}\left(\mathbf{s}, \widehat{\mathbf{s}}\right)/ \left(\omega_{CF}(\mathbf{s}, \widehat{\mathbf{s}})\times \omega_{SF}(\mathbf{s}, \widehat{\mathbf{s}})\right) 
\end{split}
\label{eq:fwdANDbkwdModelLoss_cotranMult}
\end{equation}

\noindent which are normalized by the sum of weights per mini-batch.

The issue is that CF and SF are derived from non-differentiable operations. Consequently, in $\text{\cotran}^{+}$ the feedback when added to the CE loss, does not play a role in the optimization process. In contrast, $\text{\cotran}^{\times}$ is theoretically more grounded as it optimizes a weighted CE loss. In Table~\ref{table:results_translation_fullAppendix}, along with all the previously reported results of Table 1 (in the main paper), we tabulate the performance of $\text{\cotran}^{+}$ and $\text{\cotran}^{\times}$ using only CF and using both CF, SF. When using CF alone, only the forward LLM is trained (there is no SF and thus, no need for b2b LLMs). As such, when using CF alone, the loss of $\text{LLM}_{f}$ for $\text{\cotran}^{+}$ in Eqn.~\ref{eq:fwdANDbkwdModelLoss_cotranPlus} is computed as $\mathcal{L}_f^{\theta_f}\left(\mathbf{t}, \widehat{\mathbf{t}}\right) = \begin{bmatrix} 
\mathcal{L}^{\theta_f}_{CE}\left(\mathbf{t}, \widehat{\mathbf{t}}\right) & 
    \omega_{CF}(\mathbf{t}, \widehat{\mathbf{t}}\,)
\end{bmatrix}\cdot 
\begin{bmatrix} 
    1-\alpha_c \\ \alpha_c
\end{bmatrix}$. And for $\text{\cotran}^{\times}$, the loss of $\text{LLM}_{f}$ in Eqn.~\ref{eq:fwdANDbkwdModelLoss_cotranMult} is computed as $\mathcal{L}_f^{\theta_f}\left(\mathbf{t}, \widehat{\mathbf{t}}\right) = \mathcal{L}^{\theta_f}_{CE}\left(\mathbf{t}, \widehat{\mathbf{t}}\right)/ \omega_{CF}(\mathbf{t}, \widehat{\mathbf{t}}\,)$, normalized by the sum of weights per mini-batch. 

However, it is observed from Table~\ref{table:results_translation_fullAppendix} that the LLMs struggle to effectively learn from feedback using either the additive or multiplicative approach. Their translation performance shows only minimal improvement over the baseline and does not match the effectiveness of RL-based schemes. Additionally, $\text{\cotran}^{+}$ is sensitive to the hyperparameters $\alpha_c$ and $\alpha_s$.

\subsection{Results: Surprising Finding w.r.t. Transpilers}
\label{appndixSec:transpilersSurprising}
In Table~\ref{table:results_translation_fullAppendix}, we expected human-written transpilers to outperform LLM-based methods, but they fall short on all listed metrics. The \texttt{java2python}~\cite{java2python} transpiler, last updated seven years ago, lacks support for the latest versions of Java and Python3. This highlights a common issue with hand-crafted rule-based systems -- they require manual updates after each major programming language version release. The commercial transpiler TSS CodeConv~\cite{tangiblesoftsoln} performs better but struggles with scenarios where specific conversion rules are not defined. It often ends up copying portions from the source-language code verbatim into the target-language translation. As a result, while these transpilers can provide a good starting point for human developers, they are not ideal for translating whole-programs and generating readily-compilable translations.

\subsection{Correlation between CF and SF}
\label{appndixSec:cfsfcorrelation}
Table~\ref{table:results_translation_fullAppendix} indicates a potential correlation between compiler feedback (CF) and symexec feedback (SF). Fine-tuning the LLM with CF improves the generation of compilable code and also results in more functionally equivalent code, even without direct SF fine-tuning. For example, in J2P translation, incorporating CF through RL+SFT slightly increases CompAcc from $96.12\%$ to $96.79\%$ compared to the baseline, but significantly raises FEqAcc from $44.52\%$ to $49.83\%$. Similarly, combining CF and SF via RL+SFT not only improves CompAcc rising from $73.63\%$ to $76.98\%$ in P2J, but also increases FEqAcc from $40.41\%$ to $48.68\%$. These findings suggest an implicit correlation between CF and SF, where enhancing code compilability also improves functional correctness, and vice-versa.

\begin{table*}[!ht]
\caption{\textbf{Code Translation Results (Full):} Performance Comparison of \cotran for Java-Python (J2P) and Python-Java (P2J) translation. (In each column, the highest value is marked in \textbf{bold}, second-highest \underline{underlined}.)}
\label{table:results_translation_fullAppendix}
\centering
\renewcommand{\arraystretch}{1.16}
\begin{adjustbox}{max width=1\textwidth}
\begin{tabular}{@{}cl|cccccc|cccccc@{}}
\toprule
{\multirow{2}{*}{Method / Tool}}   &   {\multirow{2}{*}{Model}}   &   \multicolumn{6}{|c}{Java $\to$ Python (J2P)}  &   \multicolumn{6}{|c}{Python $\to$ Java (P2J)} \\
\cmidrule(lr){3-8} \cmidrule(lr){9-14}
   &     &   \textit{FEqAcc}   &   \textit{CompAcc}   &   $\text{errPos}_{\nth{1}}$   &   CodeBLEU   &   BLEU   &   EM   &   \textit{FEqAcc}   &   \textit{CompAcc}   &   $\text{errPos}_{\nth{1}}$  &   CodeBLEU   &   BLEU   &   EM \\
\cmidrule(lr){1-14}
{\multirow{3}{*}{\phantom{aaaa}Transpilers}}   &   java2python~\cite{java2python}   &  3.32  &  41.46  &  28.62  &  20.31  &  17.54  &  0  &   -   &   -   &   -  &   -   &   -   &   - \\
  &   TSS CodeConv~\cite{tangiblesoftsoln}   &  0.46  &  58.30  &  54.26  &  41.87  &  24.44  &  0  &   -   &   -   &   -  &   -   &   -   &   - \\
  &   py2java~\cite{py2java}   &   -   &   -   &   -   &   -   &   -   &   -   &  0  &  0  &  1.61  &  41.56  &  48.59  &  0\\
\noalign{\vskip 0.2ex}\hdashline\noalign{\vskip 0.4ex}
{\multirow{3}{*}{\makecell{Recent \\competing tools\\\textit{(unsupervised training)}}}}   &   TransCoder~\cite{roziere2020unsupervised}   &  0.46  &  88.09  &  63.57  &  35.07  &  32.07  &  0  &  0  &  0  &  4.57  &  35.02  &  35.06  &  0\\
  &   TransCoder-DOBF~\cite{lachaux2021dobf}   &  0.46  &  63.00  &  47.10  &  39.98  &  33.84  &  0  &  0  &  0  &  3.11  &  33.33  &  32.72  &  0\\
  &   TransCoder-ST~\cite{roziere2022leveraging}   &  0.46  &  91.58  &  74.68  &  40.04  &  37.30  &  0  &  0  &  0  &  4.67  &  29.88  &  28.15  &  0\\
\noalign{\vskip 0.2ex}\hdashline\noalign{\vskip 0.4ex}
\phantom{aaaa}ChatGPT   &   \makecell[l]{GPT-3.5-turbo~\cite{chatgpt}}   &   {\textbf{76.06}}   &   {95.36}   &   {90.88}    &   {52.11}   &   {53.19}   &   {0.29}   &   {21.65}   &   {24.97}   &   {30.86}    &   {54.08}   &   {55.58}   &   {0} \\
\noalign{\vskip 0.2ex}\hdashline\noalign{\vskip 0.4ex}
{\multirow{7}{*}{\makecell{Recent \\competing tools\\(\textit{supervised training}\\\textit{on AVATAR-TC)}}}}   &   CodeBERT~\cite{feng2020codebert}   &  12.31  &  84.77  &  79.57  &  46.00  &  48.10  &  0.46  &  0.74  &   \textbf{96.79}   &   \textbf{99.51}   &  26.10  &  19.62  &  0\\
  &   GraphCodeBERT~\cite{guo2020graphcodebert}   &  10.88  &  85.05  &  79.78  &  45.53  &  47.26  &  0.57  &  0.46  &  \underline{89.75}  &  \underline{98.05}  &  23.72  &  16.21  &  0\\
  &   CodeGPT~\cite{lu2021codexglue}   &  24.86  &  78.92  &  89.21  &  38.38  &  38.64  &  1.49  &  13.40  &  45.13  &  94.50  &  40.51  &  37.96  &  0.52\\
  &   CodeGPT-adapted~\cite{lu2021codexglue}   &  24.17  &  76.75  &  89.31  &  36.84  &  37.36  &  1.55  &  20.50  &  52.00  &  97.60  &  41.46  &  38.15  &  1.03\\
  &   PLBART-base~\cite{ahmad2021unified}   &  38.55  &  91.47  &  90.79  &  54.77  &  59.34  &  1.32  &  38.26  &  75.77  &  96.64  &  55.96  &  59.24  &  0.97\\
  &   CodeT5-base~\cite{wang2021codet5}   &  40.95  &  92.84  &  \textbf{93.76}  &  55.34  &   {60.03}   &  2.41  &  33.79  &  68.84  &  98.02  &  57.64  &  60.16  &  0.86\\
  &   PPOCoder~\cite{shojaee2023execution}   &  44.27  &  93.47  &  91.44  &  55.16  &  59.51  &  1.89  &   {37.11}   &   {59.62}   &   {96.77}    &   {55.04}   &   {58.52}   &   {0.52} \\
\cmidrule(lr){1-14}
\phantom{aaaa}Our tool   &   \makecell[l]{\textbf{\cotran (baseline)}}   &  44.52  &  96.12  &  92.07  &  55.44  &  58.71  &  2.11  &  40.41  &  73.63  &  92.16  &   \textbf{59.11}   &   {61.12}   &  \underline{1.66}\\
\noalign{\vskip 0.2ex}\hdashline\noalign{\vskip 0.4ex}
{\multirow{5}{*}{\makecell{Our tool with \\compiler feedback\\only}}}   &   $\text{\cotran}^{+}$ ($\alpha_c=0.25$)   &  46.06  &  \underline{97.08}  &  93.26  &  55.18  &  58.48  &   \textbf{2.68}   &  39.95  &  74.89  &  93.94  &  58.85  &  60.64  &   \textbf{1.71} \\
  &   $\text{\cotran}^{+}$ ($\alpha_c=0.50$)   &  44.52  &  96.35  &  91.95  &  55.86  &  58.27  &  2.17  &  40.64  &  73.57  &  91.59  &  58.58  &  60.42  &  1.37\\
  &   $\text{\cotran}^{+}$ ($\alpha_c=0.75$)   &  46.33  &  96.62  &  92.89  &  55.46  &  59.68  &  \underline{2.64}  &  40.01  &  74.26  &  90.66  &  57.80  &  59.48  &  0.91\\
  &   $\text{\cotran}^{\times}$   &  44.62  &  96.91  &  \underline{93.39}  &  55.22  &  58.50  &  2.35  &  38.03  &  71.48  &  90.69  &  54.93  &  59.82  &  1.60\\
  &   \makecell[l]{\cotran + CF (RL-based fine-tuning)}   &   {47.02}   &   {96.56}   &   {91.58}    &   {56.10}   &   {60.59}   &   {2.23}   &   {42.78}   &   {74.80}   &   {96.91}    &   {58.55}   &   {61.26}   &   {1.60} \\
  &   \makecell[l]{\textbf{\cotran + CF (RL+SFT interleaved)}}   &   {49.83}   &   {96.79}   &   {92.08}    &   {56.07}   &   \underline{60.61}   &   {2.23}   &   \underline{45.93}   &   {75.77}   &   {96.89}    &   {58.28}   &   {61.21}   &   {1.60} \\
\noalign{\vskip 0.2ex}\hdashline\noalign{\vskip 0.4ex}
{\multirow{4}{*}{\makecell{Our tool (b2b) with \\compiler and symexec \\feedback}}}   &   \makecell[l]{$\text{\cotran}^{+}$ ($\alpha_c=0.5$, $\alpha_s=0.01$)}   &  45.68  &   96.58   &  92.11  &  55.72  &  58.92  &  2.17  &  40.92  &  75.27  &  94.89  &  \underline{58.97}  &  60.34  &  1.08\\
  &   \makecell[l]{$\text{\cotran}^{+}$ ($\alpha_c=0.5$, $\alpha_s=0.5$)}   &  46.14  &  96.29  &   92.02    &   55.95   &  58.22  &  2.23  &   {41.84}   &  74.89  &  93.94  &  58.89  &  60.10  &  1.31\\
  &   \makecell[l]{$\text{\cotran}^{\times}$}   &   {46.45}   &   {96.56}   &   {91.58}    &   {56.08}   &   {60.58}   &   {2.23}   &   {41.63}   &   {74.34}   &   {96.91}    &   {58.53}   &   \underline{61.27}   &   {1.60} \\
  &   \makecell[l]{\cotran + CF + SF (RL-based fine-tuning)}   &   {50.45}   &   {96.79}   &   {92.15}    &   \underline{56.17}   &   {60.60}   &   {2.23}   &   {43.92}   &   {75.14}   &   {96.93}    &   {58.59}   &   \textbf{61.28}   &   {1.60} \\
  &   \makecell[l]{\textbf{\cotran + CF + SF (RL+SFT interleaved)}}   &   \underline{53.89}   &   {\textbf{97.14}}   &   {92.73}    &   {\textbf{56.24}}   &   {\textbf{60.69}}   &   {2.29}   &   \textbf{48.68}   &   {76.98}   &   {96.93}    &   {58.38}   &   {61.19}   &   {1.60} \\
\bottomrule
\end{tabular}
\end{adjustbox}

\end{table*}

\end{document}